\documentclass[unsortedaddress,onecolumn, nofootinbib,10pt]{revtex4}
\usepackage[a4paper,bindingoffset=0.2in, left=0.5in,right=0.8in,top=1in,bottom=1in, footskip=.25in]{geometry}
\usepackage[utf8]{inputenc}
\usepackage{amsmath,empheq,color}
\usepackage{amsfonts}
\usepackage{amsthm}
\usepackage{amssymb}
\usepackage{graphicx}
\usepackage{hyperref}
\usepackage[normalem]{ulem}
\usepackage{epigraph}
\usepackage{mathrsfs}
\usepackage{bbm}
\usepackage{wrapfig}
\usepackage{lineno}
\usepackage{scalerel,stackengine}
\usepackage[T1]{fontenc}
\usepackage{epigraph}
\usepackage{cancel}
\usepackage{soul}
\usepackage{ulem}

\usepackage[textwidth=1.7cm,textsize=small]{todonotes}

\newcommand{\ncd}{\newcommand}
	\newcommand{\vsp}{\vspace{0.4cm}}

	\ncd{\mrm}    {\mathrm}
	\ncd{\beq} {\begin{equation}}
	\ncd{\eeq} {\end{equation}}

	\def\d{{\rm d}}
	\def\D{{\rm D}}
	\def\i{{\rm i}}
	
	\def\H{{\mathscr{H}}}

	\def\C{{\mathbb{C}}}
	\def\R{{\mathbb{R}}}
	\def\CP{{\mathbb{CP}}}

	\def\h{{\mathcal{H}}}

	\def\A{{\mathcal{A}}}

	\def\F{{\mathfrak{F}}}
	\def\X{{\mathfrak{X}}}

	\def\a{{\mathfrak{A}}}
	\def\D{{\mathfrak{D}}}

	\def\e{{\rm e}}
	
	\def\Tr{{\rm Tr}}
	
\begin{document}

\title{From geometry to coherent dissipative dynamics in quantum mechanics}

\author{Hans Cruz-Prado}
\email{hans@ciencias.unam.mx}
\affiliation{Instituto de Ciencias Nucleares, Universidad Nacional Aut\'onoma de M\'exico, Apartado Postal 70543, Ciudad de M\'exico 04510, M\'exico.}

\author{Alessandro Bravetti}
\email{alessandro.bravetti@iimas.unam.mx} 
\affiliation{Instituto de Investigaciones en Matem\'aticas Aplicadas y en Sistemas, Universidad Nacional Aut\'onoma de M\'exico, A. P. 70543, M\'exico, DF 04510, Mexico}

\author{Angel Garcia-Chung}
\email{alechung@xanum.uam.mx} 
\affiliation{Departamento de F\'isica, Universidad Aut\'onoma Metropolitana - Iztapalapa, \\ San Rafael Atlixco 186, Ciudad de M\'exico 09340, M\'exico.}
\affiliation{Tecnol\'ogico de Monterrey, Escuela de Ingenier\'ia y Ciencias, Carr. al Lago de Guadalupe Km. 3.5, Estado de Mexico 52926, Mexico.}

\begin{abstract}
Starting from the geometric description of quantum systems, we propose a novel approach to time-independet 
dissipative quantum processes according to which the energy is dissipated but the coherence of the states is preserved.
Our proposal consists on extending the standard symplectic picture of quantum mechanics to a contact manifold
and then obtaining dissipation using an appropriate contact Hamiltonian dynamics.
We work out the case of finite-level systems, for which it is shown by means of the corresponding
 contact master equation that the resulting dynamics constitutes 
a viable alternative candidate for the description of this subclass of dissipative quantum systems.
As a concrete application, motivated by recent experimental observations, we describe quantum decays in a $2$-level system 
as coherent and continuous processes. 
\end{abstract}

\maketitle


\section{Motivation and previous works}

Dissipative quantum phenomena have been the subject of intense investigation since the early days of quantum mechanics~\cite{landau1927}.
The most widely adopted description of dissipative quantum processes is given by  
the \emph{Gorini--Kossakowski--Sudarshan--Lindblad (GKLS) equation}~\cite{gorini1976completely,lindblad1976generators,chruscinski2017brief}.
This is because it has been proven that the GKLS equation describes the most general form of a non-unitary, linear, completely positive and trace-preserving dynamics; 
thus, one may ensure that the probabilistic
character of quantum mechanics is preserved at all times.
However, in the GKLS equation in order to have a completely positive evolution it is necessary to add the so-called \emph{jump term}, which changes the rank of the state, and it is therefore responsible for decoherence. 
As a consequence, the GKLS dynamics fundamentally couples energy dissipation and decoherence.

On the other hand, it is posible to model energy dissipation without decoherence by means of a time-dependent Hamiltonian, 
where the time dependence describes in an effective way the interaction of the system with the environment; 
in this case, the evolution is unitary and therefore the rank of the state is preserved. 
For instance, recent impressive improvements in experimental settings have allowed to observe  
the decays induced by a Rabi oscillator in a $2$-level system as coherent and continuous processes~\cite{minev2019catch} (see also~\cite{snizhko2020quantum} for a theoretical analysis).

In this work, we propose an alternative way to describe a class of  time-independent dissipative quantum phenomena  by a non-unitary evolution
without introducing decoherence. 
This is based on the symplectic formulation of quantum mechanics and on the analogy with the geometric description of classical dissipative systems: 
in classical mechanics one can describe a wide class of dissipative systems by recurring to the \emph{contactification} 
of the symplectic phase space and then using contact Hamiltonian systems to define the dynamics. 
It has been shown that this approach, when applicable, provides several positive features, 
such as the fact of relying on canonical variables and producing a generalization of canonical transformations~\cite{bravetti2017contact}, 
enabling an extension of both Liouville and Noether's theorems to 
the dissipative case~\cite{bravetti2015liouville,lazo2019noether,gaset2020new,de2020infinitesimal,bravetti2020invariant,bravetti2021geometric},
and providing a description in terms of variational principles~\cite{georgieva2002first,georgieva2003generalized,wang2016implicit,liu2018contact,wang2019aubry,vermeeren2019contact,cannarsa2019herglotz},
together with a natural route to field theories with dissipation~\cite{gaset2020contact}.

Analogously, in the geometric description of $n$-level quantum systems, the pure states of the corresponding
Hilbert space $\h$ are points on the complex projective space $\CP(\h_{0})$, 
which is a symplectic manifold, 
and the Schr\"odinger dynamics on $\h$ is projected onto a 
Hamiltonian dynamics on such manifold~\cite{ashtekar1999geometrical, brody2001geometric, ercolessi2010equations, cruz2020nonlinear}.
This is because pure quantum states in $\h$ are actually \emph{rays}
and the Schr\"odinger equation is invariant under dilations and multiplications by a global phase factor, that is, in the standard Schr\"odinger picture one deals with redundant information in order to obtain a linear description.
Interestingly, contact manifolds also appear naturally in this picture. In fact, by quotienting only over dilations 
one obtains the manifold of normalized vectors in $\h$ identified as the $(2n-1)$-dimensional unit sphere $S^{2n-1}(\h)$ with the standard contact structure.
An approach to describe dissipation in the quantum case by means of contact Hamiltonian flows on $S^{2n-1}(\h)$ has been put forward already in~\cite{ciaglia2018contact}.

Contrary to~\cite{ciaglia2018contact}, we do not work on $S^{2n-1}(\h)$, but we proceed by analogy with the classical case, i.e., we perform a contactification of $\CP(\h_{0})$ and investigate the dissipative dynamics by means of contact Hamiltonian systems defined on the extended space $\CP(\h_{0})\times\mathbb R$.
The reason is the fact that on $S^{2n-1}(\h)$ the global phase of the vector is still present and therefore the description still depends on an unphysical degree of freedom, which may enter non-trivially the dynamics for the dissipative case.
Hence, we consider here the more conservative approach of working directly with the physical states, i.e.,~points on $\CP(\h_{0})$, and defining a dissipative dynamics directly on such space.

The structure of this work is as follows: after a brief review of the standard geometric description of $n$-level quantum systems in Section~\ref{sec:GQM}, we introduce our approach in Section~\ref{sec:CHCP} and we
show that, by choosing the contact Hamiltonian appropriately, the dynamics on $\CP(\h_{0})\times\mathbb R$ is projectable onto a proper dynamics on $\CP(\h_{0})$, thus preserving the purity of states, while at the same time dissipating the expected value of the energy of the reference system. 
In this way we obtain a dissipative dynamics on the manifold of physical quantum states.
Then, in Section~\ref{sec:CHDM} we consider the corresponding dynamics for the density operators, which we call \emph{the contact master equation} and, after a brief comparison with other approaches, we prove that the resulting map is both positive and trace preserving, thus agreeing with all the prescriptions of standard quantum mechanics.
As a consequence, the proposed contact dynamics is a viable candidate to describe dissipative phenomena in quantum systems when the purity of the states is preserved.
As an illustration of our formalism, we consider in Section~\ref{sec:Example} the important case of radiative decay in qubit systems, finding that our description can effectively model quantum decays (or excitations) as coherent and continuous processes.
Finally, in Section~\ref{sec:Conclusions} we summarize our results and highlight future directions.


\section{Geometry and dynamics of conservative $n$-level quantum systems}\label{sec:GQM}

Let us start our study by recalling some aspects of the geometric description of finite-level quantum systems. 
For complete reviews we refer to~\cite{ashtekar1999geometrical,brody2001geometric,ercolessi2010equations,carinena2015geometry,bengtsson2017geometry},
while further results can be found in~\cite{carinena2017tensorial,ciaglia2017dynamical,ciaglia2018contact,cruz2020nonlinear}.


\subsection{Kinematics: From $\h_0$ to $\CP(\h_0)$ }\label{subsec:Skinematics}

It is well-known that in the Hilbert space there is a natural action of the Abelian group $\mathbb{C}_{0} = \C - \{\, 0 \, \}$ given by
	\beq	
	|\psi\rangle\;\mapsto \lambda\,|\psi\rangle\,=\,\varrho\,\mathrm{e}^{\i \, \theta}\,|\psi\rangle\,
	\quad
	\mbox{with}
	\quad	
	\varrho>0 \, ,
	\eeq
which implies that a pure state is an equivalence class, i.e.~\emph{a ray} in the Hilbert space $\h$.
Therefore complete measurements in quantum mechanics yield an equivalence class of vectors. 

Let us show how, for finite-dimensional systems, the set of such equivalence classes can be given the structure of a manifold:
 consider an $n$-level quantum system with Hilbert space $\h$ and define $\h_0 = \h - \{ \mathbf{0} \}$.
Selecting an orthonormal basis $\{ | e_k \rangle \}_{k = 1, \dots, n}$ in $\h_{0}$, one may introduce a Cartesian coordinate system $\{ x^k , y^k \}_{k = 1, \dots, n}$ on $\h_0$, such that for any element $| \psi \rangle \in \h_0$ one has that
	\beq \label{coord-sys}
	| \psi \rangle = \psi^k | e_k \rangle = (x^k + \i \, y^k) | e_k \rangle \, ,
	\eeq
where here and in the following equations Einstein's summation convention over repeated indices is assumed. 
The action defining the equivalence classes of pure quantum states 
may be described infinitesimally by means of two commuting linear vector fields, given in Cartesian coordinates as
	\beq\label{eqn: action on the hilbert space generating the sphere and the complex projective space}
	\Delta=\, x^{k}\,\frac{\partial}{\partial x^{k}} + y^{k}\,\frac{\partial}{\partial y^{k}}
	\quad
	{\rm{and}}
	\quad
	\Gamma\,=y^{k}\,\frac{\partial}{\partial x^{k}} - \,x^{k}\,\frac{\partial}{\partial y^{k}} \, ,
	\eeq
where $\Delta$ is the infinitesimal generator of dilations, 
while $\Gamma$ is the infinitesimal generator of the multiplication by a global phase factor.
Now, dilations define a distribution 
whose integral curves 
foliate the Hilbert space.
Let $\Phi^\Delta$ denote such foliation, then the quotient space $\h_{0}/\Phi^\Delta$ is \emph{the unit sphere}
	\beq
	S^{2n-1}(\h)\,:=\,\left\{|\psi\rangle \in \h_0 \; | \, \langle\psi|\psi\rangle\,=\,1\right\}.
	\eeq
In the following an element of $S^{2n-1}$ is denoted as $|\psi)$, whereas $|\psi\rangle$ is a vector in $\h_{0}$.
Now let $\Phi^{\Delta,\Gamma}$ be the foliation corresponding to the distribution generated by $\Delta$ and $\Gamma$.
Then the quotient space $\h_{0}/\Phi^{\Delta,\Gamma}$ is identified with \emph{the complex projective space} $\mathbb{CP}(\h_0)$, defined as
	\beq
	\mathbb{CP}(\h_0)\, := \, \{ \lambda \, | \psi \rangle | \, \lambda \in \C_0 \,  \} \, .
	\eeq
Therefore, points $[\psi] \in \CP(\h_0)$ are identified with the pure states of the quantum system.

Furthermore, there is a one-to-one correspondence between elements of $\CP(\h_0)$ and rank-one projectors, given by
	\beq \label{st-mat}
	[\psi] \mapsto \rho_\psi : =  \frac{| \psi \rangle \langle \psi |}{\langle \psi | \psi \rangle} \, .
	\eeq

On $\CP(\h_0)$ it is convenient to work with \emph{complex homogeneous coordinates}. To introduce such coordinates,  
let $U_j \subset \CP(\h_0)$ denote the open subset where $\psi^{j}\neq 0$. 
Then on $U_{j}$ one may introduce the coordinates 
	\beq \label{CP-H}
	\phi_j : U_j \to \C^{n-1}: 
	[\psi^1, \dots, \psi^n] \mapsto \left( z^1, \dots, z^{j-1},
	z^{j + 1}, \dots , z^n \right)\, ,
	\quad
	\text{with}
	\quad
	z^k = \frac{\psi^k}{\psi^j} \, .
	\eeq  
The set of $(U_j, \phi_j)$, with $j=1, \dots, n$, constitutes an \emph{atlas} for $\CP(\h_0)$.
In such coordinates, the projection from $\h_0$ onto $\CP(\h_0)$ can be explicitly expressed as
	\beq \label{Riccati-proj}
	\pi : \h_0 \to \CP(\h_0) : | \psi \rangle \mapsto |\psi] = \frac{1}{\sqrt{1 + | \mathbf{z} |^2 } }
	\left(
	\begin{array}{c}
	\mathbf{z} \\
	1 
	\end{array}
	\right)  \, ,
	\eeq
where for simplicity we used the chart $\phi_n$  and we introduced the notations
	\begin{equation}
	\mathbf{z} = \left( \begin{array}{c} z^1 \\ z^2 \\ \vdots \\ z^{n-1} \end{array}\right) \qquad \text{and} \qquad | \mathbf{z} |^2 =  \mathbf{z}^\dagger \, \mathbf{z}\,.
	\end{equation}

To conclude the kinematical analysis of $n$-level systems, let us recall that
$\CP(\h_0)$ is a K\"ahler manifold~\cite{ercolessi2010equations, carinena2015geometry, mcduff2017};
in fact, considering the homogeneous coordinates in Eq.~\eqref{Riccati-proj}, one can introduce the $1$-form 
	\begin{equation}\label{1-form}
	\theta_{\tiny \mbox{FS}} 
	 = \frac{\hbar}{\i} \frac{1}{\sqrt{1 + | \mathbf{z} |^2}}
	( \, \bar{z}_k \, , \, 1 \, )
	\left[ \frac{1}{\sqrt{1 + | \mathbf{z} |^2}} 
	\left(
	\begin{array}{c}
	  \d z^k \\
	  0
	\end{array}
	\right)
	  + 
	\left(
	\begin{array}{c}
	  z^k  \\
	  1
	\end{array}
	\right)  
	\d\left(  \frac{1}{\sqrt{1 + | \mathbf{z} |^2}} \right) \right] 
	 =  \frac{\hbar}{2 \, \i } \frac{\bar{z}_k \, \d z^k - z^k \, \d \bar{z}_k }{1 +  | \mathbf{z} |^2 } \, .
	\end{equation}
Then the symplectic form $\omega_{\tiny \mbox{FS}}$ on $\CP(\h_0)$ is given by the exterior derivative $\omega_{\tiny \mbox{FS}} = \d \theta_{\tiny \mbox{FS}}$,
which reads
	\begin{equation} \label{Sym-form}
	\omega_{\tiny \mbox{FS}} = \frac{- \i \, \hbar}{(1 + | \mathbf{z} |^2 )^2} 
	\left[ 
	(1 + | \mathbf{z} |^2 ) \, \d \bar{z}_k \wedge \d z^k 
	- \frac{1}{2} \,  ( \bar{z}_l \, \d z^l + z^l \, \d \bar{z}_l) \wedge  ( \bar{z}_k \, \d z^k - z^k \, \d \bar{z}_k )
	\right] \, .
	\end{equation}

Moreover, 
as proven by Wootters in~\cite{wootters1980}, a natural notion of distance between quantum states is given by the Fubini-Study metric on $\CP(\h_{0})$, which has the form~\cite{ercolessi2010equations, carinena2015geometry,bengtsson2017geometry, mcduff2017}

	\begin{equation} \label{Grad-form}
	g_{\tiny \mbox{FS}} 
	 = \frac{ - \hbar}{(1 + | \mathbf{z} |^2 )^2} 
	\Bigg[ 
	(1 + | \mathbf{z} |^2) \, \d \bar{z}_k \otimes_{\tiny \mbox{S}} \d z^k 
	+ \frac{1}{2} \, ( \bar{z}_k \, \d z^k - z^k \, \d \bar{z}_k ) \otimes ( \bar{z}_l \, \d z^l - z^l \, \d \bar{z}_l ) 			
	 - \frac{1}{2} \,  ( \bar{z}_k \, \d z^k + z^k \, \d \bar{z}_k)  \otimes  ( \bar{z}_l \, \d z^l + z^l \, \d \bar{z}_l ) 
	\Bigg] \, ,
	\end{equation}
where $\d \bar{z}_k \otimes_{\tiny \mbox{S}} \d z^k = \d \bar{z}_k \, \d z^k + \d z^k \, \d \bar{z}_k$. 
On the other hand, the  $(1,1)$-tensor 
	\beq \label{Com-st-FS}
	J_{\tiny \mbox{FS}} = \frac{1}{\i}
	\left(
	\d z^k \otimes \frac{\partial}{\partial z^k} - \d  \bar{z}_k \otimes \frac{\partial}{\partial  \bar{z}_k }
	\right) \, 
	\eeq
defines a complex structure on $\CP(\h_0)$, such that the quadruple
$(\CP(\h_0),\omega_{\tiny \mbox{FS}}, g_{\tiny \mbox{FS}}, J_{\tiny \mbox{FS}})$ is a 
K\"ahler manifold~\cite{ercolessi2010equations, carinena2015geometry,bengtsson2017geometry, mcduff2017}.


\subsection{Dynamics: symplectic Hamiltonian systems on $\CP(\h_0)$}\label{subsec:SHDCPh}

Having defined the canonical projection $\pi$ from $\h_0$ to $\CP(\h_0)$, one may project the dynamics of the system
as follows:
first we recall that the Schr\"odinger dynamics may be viewed as a classical Hamiltonian  system on $\h_{0} \times \h_{0}$ 
with the symplectic structure
given by the imaginary part of the Hermitian scalar product
and the generating Hamiltonian function being the expectation value $e_\mathbf{H}=\langle\psi|\mathbf{H}|\psi\rangle$ of the Hamiltonian operator $\mathbf{H}$.
The corresponding Hamiltonian vector field then reads
 	\beq
	X_\mathbf{H} = \frac{\i}{\hbar} \, \bar{\psi}_j \, H^j_k \, \frac{\partial }{\partial \bar{\psi}_k}
	- \frac{\i}{\hbar} \, H^k_j \, \psi^j \frac{\partial }{\partial \psi^k} \, ,
	\eeq
where $H^j_k$ are the entries of $\mathbf{H}$~\cite{ercolessi2010equations}. 
Now, because $[X_\mathbf{H}, \Delta] = [X_\mathbf{H}, \Gamma] = 0$, i.e.~$\Delta$ and $\Gamma$  
are symmetries of the Schr\"odinger dynamics, it is posible to project $X_{\mathbf{H}}$ onto a dynamics on $\CP(\h_0)$~\cite{marmo1985}.
It turns out that the projected dynamics $X_{e_\mathbf{H}}$
is again a Hamiltonian vector field with respect to the symplectic structure $\omega_{\tiny \mbox{FS}}$,
with generating Hamiltonian function $e_\mathbf{H} \in \F(\CP(\h_0))$, that is, it satisfies
	\beq \label{Met-Sym-FS}
	\omega_{\tiny \mbox{FS}}(X_{e_\mathbf{H}}, \, \cdot \,)  = \d \, e_\mathbf{H} \,.
	\eeq 

In order to express Eq.~\eqref{Met-Sym-FS} in complex homogeneous coordinates on $\CP(\h_0)$, 
we start by writing the expectation value of the observable $\mathbf{H}$ in such coordinates as
	\begin{equation} \label{Exp-Val-Ham}
	e_\mathbf{H} 
	 =  [ \psi | \mathbf{H} | \psi ] 
	 = 
	\frac{1}{1 + | \mathbf{z} |^2 } 
	\, \,
	( \,  \mathbf{z}^\dagger  \,\,\, , \,\,\, 1 \, )
	\left(
	\begin{array}{ccc}
	\mathbb{H}_1  &  \mathbb{V}   \\
	& \\
	\mathbb{V}^\dagger &  H_2   \\   
	\end{array}
	\right)
	\left(
	\begin{array}{c}
	 \mathbf{z} \\
	 \\
	 1
	\end{array}
	\right) 
	 =  \,
	\frac{1}{1 + | \mathbf{z} |^2 } \left( \, 
	\mathbf{z}^\dagger \mathbb{H}_1 \mathbf{z} + \mathbf{z}^\dagger \mathbb{V} 
	+ \mathbb{V}^\dagger \mathbf{z} + H_2
	\, \right) \, ,
	\end{equation}
where $\mathbb{H}_1$ is an $(n-1) \times (n-1)$--dimensional matrix, $\mathbb{V}$ is an $(n-1)$--dimensional column vector and $H_2$ a real quantity. 
Thus, using~\eqref{Sym-form} and~\eqref{Met-Sym-FS}, 
we get
	\beq \label{Ricc-dyn}
	X_{e_\mathbf{H}} = X_{z^k} \, \frac{\partial}{\partial z^k}
	+ X_{\bar{z}_k} \, \frac{\partial}{\partial \bar{z}_k} \, , 
	\eeq
where the component $X_{\bar{z}_k}$ is the complex conjugate of $X_{z^k}$, and 
	\beq \label{N-Ham-Vec}
	X_{z^k} = \frac{\i}{\hbar} 
	\left(
	z^k \, \bar{V}_l \, z^l - |\mathbb{H}_1|^k_l \, z^l + H_2 \, z^k - V^k 
	\right) \, .
	\eeq
Therefore, the integral curves of this Hamiltonian vector field are solutions to the Hamiltonian equations of motion
	\beq \label{N-Ev}
	\dot{z}^k = - \frac{\i}{\hbar} \, (1 + | \mathbf{z} |^2 ) 
	\left( 
	\frac{\partial e_\mathbf{H}}{\partial \bar{z}_k} 
	 +
	  z^k \, \bar{z}_l \, \frac{\partial e_\mathbf{H}}{\partial \bar{z}_l} 
	  \right) 
	 =  \frac{\i}{\hbar}
	\left(
	z^k \, \bar{V}_l \, z^l - |\mathbb{H}_1|^k_l \, z^l + H_2 \, z^k - V^k 
	\right)  \, , 
	\eeq
which is also called the \emph{matrix Riccati equation}~\cite{chaturvedi2007ray}.
Hence, the matrix Riccati equation is simply the coordinate expression of the projection of the Schr\"odinger equation onto the complex projective space.

In addition, one may use the symplectic structure to introduce a Poisson bracket on $\CP(\h_{0})$~\cite{ercolessi2010equations, carinena2015geometry}:
given the expectation values $e_\mathbf{A}$ and $e_\mathbf{B}$ associated with the observables $\mathbf{A}$ and $\mathbf{B}$, one defines
	\beq
	\{ e_\mathbf{A}, e_\mathbf{B} \}_{\omega_{\tiny \mbox{FS}}} 
	= \omega_{\tiny \mbox{FS}}(X_{e_\mathbf{A}}, X_{e_\mathbf{B}} ) \,.
	\eeq
In complex homogeneous coordinates this bracket reads
	\beq \label{Exp-Pois-Bra}
	\{ e_\mathbf{A}, e_\mathbf{B} \}_{\omega_{\tiny \mbox{FS}}} 
	= - \frac{\i}{\hbar} (1 +   | \mathbf{z} |^2 ) 
	\left[
	\left(
	\frac{\partial e_\mathbf{A}}{\partial z^k} \, \frac{\partial e_\mathbf{B}}{\partial \bar{z}_k} 
	- \frac{\partial e_\mathbf{A}}{\partial \bar{z}_k} \, \frac{\partial e_\mathbf{B}}{\partial z^k} 
	\right)
	+
	\left( 
	z^k \, \frac{\partial e_\mathbf{A}}{\partial z^k} \, \bar{z}_l \frac{\partial e_\mathbf{B}}{\partial \bar{z}_l} 
	-
	\bar{z}_l \, \frac{\partial e_\mathbf{A}}{\partial \bar{z}_l} \, z^k \frac{\partial e_\mathbf{B}}{\partial z^k} 
	\right)
	\right] \, .
	\eeq
Furthermore, after some calculations it is possible to prove that the Poisson bracket satisfies 
$\{ e_\mathbf{A} , e_\mathbf{B} \}_{\omega_{\tiny \mbox{FS}}} ~=~e_{\frac{1}{\i \, \hbar}[\mathbf{A} , \mathbf{B} ]}$
where $[\mathbf{A} , \mathbf{B} ] = \mathbf{A} \, \mathbf{B} - \mathbf{B} \, \mathbf{A}$.
Therefore, one has a clear connection between the Poisson bracket and the quantum commutator. 
In particular, considering the Hamiltonian of the system $\mathbf{H}$ with expectation value $e_\mathbf{H}$, then the evolution of the expectation value $e_\mathbf{A}$ of an arbitrary observable $\mathbf{A}$ is given by
	\beq
	\frac{\d e_\mathbf{A}}{\d t} = \{ e_\mathbf{A}, e_\mathbf{H} \}_{\omega_{\tiny \mbox{FS}}} 
	= e_{\frac{1}{\i \, \hbar}\, [\mathbf{A}, \mathbf{H} ]}\, .
	\eeq  
This result implies immediately that in the time-independent case $e_\mathbf{H}$ is a first integral of the flow, i.e., 
the expectation value of the Hamiltonian is conserved. In addition, the expectation value of any observable commuting with $\mathbf{H}$ is a first integral too.

Proceeding in parallel with the above construction of Hamiltonian vector fields and of the Poisson
bracket by means of the symplectic structure, we are going to define now gradient vector fields $Y_{e_\mathbf{H}}$ 
and the Jordan bracket using  the Fubini--Study metric.

Gradient vector fields are defined as
	\beq
	g_{\tiny \mbox{FS}}( Y_{e_\mathbf{H}} , \, \cdot \,)  = \d \, e_\mathbf{H} \, .
	\eeq
and it is direct to verify that 
	\beq\label{eq:relJ}
	J_{\tiny \mbox{FS}} (X_{e_\mathbf{H}}) = Y_{e_\mathbf{H}}\,,
	\eeq 
with $J_{\tiny \mbox{FS}}$ given in~\eqref{Com-st-FS}.
Therefore, employing Eqs.~\eqref{Ricc-dyn} and~\eqref{N-Ham-Vec} one finds the explicit coordinate expression for  the gradient vector field,
which reads
 	\begin{align} \label{Symp-Grad}
	Y_{e_\mathbf{H}} 
	& 
	=  \frac{1}{\hbar} \left[ z^k \, \bar{V}_l \, z^l - |\mathbb{H}_1|^k_l \, z^l + H_2 \, z^k - V^k \right] \, 
	\frac{\partial}{\partial z^k}
	+ \frac{1}{\hbar} \left[ \bar{z}_l \, V^l \, \bar{z}_k - \bar{z}_l \,  |\mathbb{H}_1|^l_k + H_2 \, \bar{z}_k - \bar{V}_k 
	 \right] \, \frac{\partial}{\partial \bar{z}_k} \, .
	\end{align}
On the other hand, the Jordan bracket between the expectation values $e_\mathbf{A}$ and $e_\mathbf{B}$ is defined as
	\beq\label{eq:Jordan}
	\{ e_\mathbf{A}, e_\mathbf{B} \}_{g_{\tiny \mbox{FS}}} 
	= g_{\tiny \mbox{FS}}(Y_{e_\mathbf{A}}, Y_{e_\mathbf{B}}) \, ,
	\eeq
and in coordinates we have
	\beq \label{Exp-Grad-Bra}
	\{ e_\mathbf{A} , e_\mathbf{B} \}_{g_{\tiny \mbox{FS}}} = - \frac{1}{\hbar} (1 +  \bar{z}_l \, z^l ) 
	\left[
	\left(
	\frac{\partial e_\mathbf{A}}{\partial z^k} \, \frac{\partial e_\mathbf{B}}{\partial \bar{z}_k} 
	+ \frac{\partial e_\mathbf{A}}{\partial \bar{z}_k} \, \frac{\partial e_\mathbf{B}}{\partial z^k} 
	\right)
	+
	\left( 
	z^k \, \frac{\partial e_\mathbf{A}}{\partial z^k} \, \bar{z}_l \frac{\partial e_\mathbf{B}}{\partial \bar{z}_l} 
	+
	\bar{z}_l \, \frac{\partial e_\mathbf{A}}{\partial \bar{z}_l} \, z^k \frac{\partial e_\mathbf{B}}{\partial z^k} 
	\right)
	\right]  \, .
	\eeq
Finally, the Jordan bracket is connected with the dispersion and the correlation of the observables. 
This is, for every couple of observables $\mathbf{A}$ and $\mathbf{B}$ their uncertainties and correlations are given by
	\beq \label{varianza}
	\sigma^2_\mathbf{A} = e_{\mathbf{A}^2} - e^2_\mathbf{A} 
	= - \frac{\hbar}{2} \{ e_\mathbf{A} , e_\mathbf{A} \}_{g_{\tiny \mbox{FS}}}
	\eeq
and
	\beq \label{correlation}
	\sigma_{\mathbf{A}\mathbf{B}} =  \frac{1}{2} \, e_{[\mathbf{A} , \mathbf{B} ]_{+}} 
	- e_{\mathbf{A}} \, e_{\mathbf{B}}
	= - \frac{\hbar}{2} \{ e_\mathbf{A} , e_\mathbf{B} \}_{g_{\tiny \mbox{FS}}} \, ,
	\eeq  	
respectively, with $[\mathbf{A} , \mathbf{B} ]_{+}=\mathbf{A}\mathbf{B}+\mathbf{B}\mathbf{A}$ being the anti-commutator. 
Thus, the Riemannian metric on $\CP(\h_{0})$ takes into account the probabilistic character of quantum mechanics~\cite{cirelli1990}. 	


\subsubsection*{Example: the conservative qubit}

As an example, let us consider a qubit, for which $\CP(\h_0)$ is 2-dimensional and each point is given by $[\psi^{1},\psi^{2}]$.
For this case the complex projective space may be thought of as the unit sphere
	\beq \label{sphere}
	S^2 = \{ (x^1, x^2, x^3) \in \R^3 \, | \, (x^1)^2 + (x^2)^2 + (x^3)^2 = 1 \} \, ,
	\eeq
and one may use the homogeneous coordinates on $\CP(\h_0)$  to induce coordinates on $S^2$ as follows:
take each coordinate patch $U_{j} \in \CP(\h_0)$ where $\psi^j \neq 0$, $j=1,2$ and introduce the homogeneous coordinates on $U_{1}$ (resp. $U_{2}$) as
defined above, that is,
	\beq \label{CP(2)-atlas}
	\phi_1 :  [\psi^1, \psi^2] \mapsto z = \frac{\psi^1}{\psi^2}
	\quad
	\left(\text{resp.}
	\quad
	\phi_2 :  [\psi^1, \psi^2] \mapsto \zeta = \frac{\psi^2}{\psi^1}\right) \,.
	\eeq
Then writing $z = z_{\tiny\mbox{R}} + \i \, z_{\tiny\mbox{I}}$
and using the stereographic projection from the north pole of the sphere one obtains the corresponding point $(x^1, x^2, x^3) \in S^2$ by
	\beq \label{Stereographic-Projection-2}
	x^1 = \frac{2 z_{\tiny\mbox{R}} } {1 + |{z }|^2} \, ,
	\qquad
	x^2 = \frac{ 2 z_{\tiny\mbox{I}} } {1 + |{z } |^2} \, ,
	\qquad
	x^3 = \frac{-1 + | {z } |^2}{1 + |{z } |^2}\,.
	\eeq

\begin{figure}[h!]
	\centering
	\includegraphics[width = 10 cm]{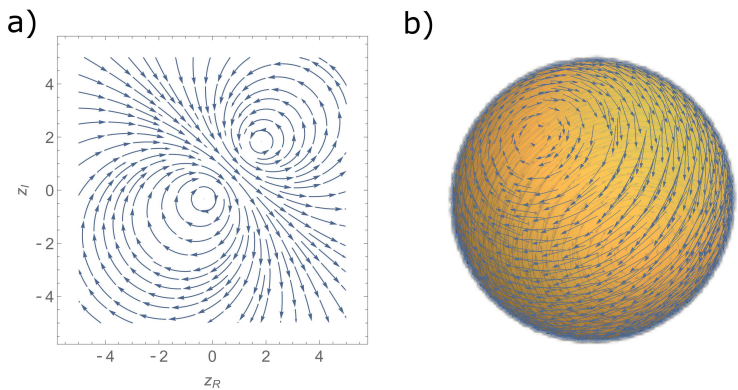}
	\caption{ a) Phase portrait of the Hamiltonian vector field (\ref{Ricc-Ham-Eq}) on the plane $\C$. b) The same vector field on the sphere $S^2$ using the map (\ref{Stereographic-Projection-2}). In both cases, the values of the parameters are $H_1=4$, $H_2=2$ and $V = 1 + \i$.}
	\label{Fig-2}
	\end{figure}

For this example let us consider the Hamiltonian operator as the matrix
	\beq
	\mathbf{H}_{\text{q}} =  \left(
	\begin{array}{ccc}
	H_1  &  V   \\
	& \\
	\overline{V} &  H_2   \\   
	\end{array}
	\right) \, ,
	\eeq 
so that the expectation value of $\mathbf{H}_{\text{q}}$ is given by 
	\begin{equation} \label{exp-value}
	e_{\mathbf{H}\text{q}} 
	 = 
	 [ \psi | \mathbf{H}_{\text{q}} | \psi ]
	 = 
	\frac{1}{1 + | {z} |^2} 
	\, \,
	( \,  \bar{z}  \,\,\, , \,\,\, 1 \, )
	\left(
	\begin{array}{ccc}
	H_1  &  V   \\
	& \\
	\overline{V} &  H_2   \\   
	\end{array}
	\right)
	\left(
	\begin{array}{c}
	 z   \\
	 \\
	 1
	\end{array}
	\right) 
	 =  \,
	\frac{1}{1 + | {z} |^2} \left[ \, 
	 H_1 | {z} |^2 + \bar{z} V + \overline{V} z + H_2
	\, \right] \,,
	\end{equation}
and Hamilton's equations~\eqref{N-Ev} reduce to
	\beq \label{Ricc-Ham-Eq}
	\dot{z} = \frac{\i}{\hbar} \left[ \overline{V} z^2 - (H_1 - H_2) z - V \right] \, .
	\eeq
To give a qualitative description of the behavior of this nonlinear system, one first realizes that the critical points of the flow are located at
	\beq \label{Cons-Sing-Point}
	z_{\pm} = \frac{(H_1 - H_2) \pm \sqrt{ (H_1 - H_2)^2 + 4 | V |^2} }{ 2 \overline{V} } \,,
	\eeq
and they are both centers. 
The phase portrait in the complex plane is depicted in Fig.~\ref{Fig-2}a, while the corresponding 
vector field on $S^2$ obtained by means of~\eqref{Stereographic-Projection-2} is displayed in Fig.~\ref{Fig-2}b. 

Another important quantity that may be easily calculated by means of~\eqref{varianza} is the uncertainty of the energy, 
given by
	\beq \label{Conservative-Unc}
	\sigma^2_{\mathbf{H}\text{q}} =  \frac{1}{(1 + | {z} |^2)^2} 
	 | V \bar{z}^2 - (H_1 - H_2) \bar{z} - \overline{V} |^2 \,.
	\eeq
Clearly this is a positive quantity that only vanishes at the singular points $z_{\pm}$.


\section{Contact geometry and dynamics of dissipative $n$-level quantum systems}\label{sec:CHCP}

In this section we consider a particular class of dissipative quantum systems, those that admit a contact Hamiltonian description
(see also \cite{bravetti2017contact,guha2017generalized,ciaglia2018contact,sloan2018dynamical,
carinena2019nonstandard,de2020infinitesimal,gaset2020new,gaset2020contact,de2020review} 
for detailed discussions on the strengths and limitations of this approach
both in the classical and quantum settings).

\subsection{Kinematics: from $\CP(\h_0)$ to $\CP(\h_0)\times \mathbb R$}\label{subsec:Contactification}

To introduce dissipation, we will work on the contactification of $\CP(\h_0)$.
Therefore, first of all let us recall some basic facts of contact geometry~\cite{MR2397738, blair2010riemannian}.

A $(2n+1)$-dimensional manifold $M$ is said to be an \emph{exact contact manifold} 
if it is endowed with a global differentiable $1$-form $\eta$  such that $\eta \wedge (\d \eta)^n \neq 0$ everywhere on $M.$ 
Then $\eta$ is called the \emph{contact form} and the \emph{contact structure} on $M$ is given by the (non--integrable)
distribution of hyperplanes $\mathcal{D}={\rm ker}(\eta)$. 
To introduce a contact manifold for finite-dimensional quantum mechanical systems, 
we perform a contactification of the space of pure quantum states $\CP(\h_0)$.
Indeed, since $\CP(\h_0)$ is an exact symplectic manifold, one can consider the extended space $\CP(\h_0) \times \R$,  
which carries a natural contact structure, given as the kernel of the contact form
	\beq \label{one-form}
	\eta = \d S  -  \theta_{\tiny \mbox{FS}} \, ,
	\eeq
where $S$ is the global coordinate on the fiber $\R$, and we use a slight abuse of notation by indicating as  $\theta_{\tiny \mbox{FS}}$ the pullback
of this 1-form on $\CP(\h_0)$ by means of the natural projection from $\CP(\h_0) \times \R$ onto the first argument.

To conclude the kinematical analysis of dissipative n-level systems we shall introduce, in analogy with the above conservative case,
some further geometric structures.
We start with the definition of a Riemmanian metric on $\CP(\h_0) \times \R$ given by
	\beq\label{eq:gFSE}
	g = g_{\tiny \mbox{FS}} + \frac{2}{\hbar} \, \eta \otimes \eta \, ,
	\eeq
\noindent where $g_{\tiny \mbox{FS}}$ is the Fubini--Study metric defined in~\eqref{Grad-form} and $\eta$ is the contact form~\eqref{one-form}.
Furthermore, $\CP(\h_0) \times \R$ admits a field $\varphi$ of endomorphisms satisfying $\varphi^2 = - I + \eta \, \otimes \, \xi$, defined by
	\beq
	\varphi = J_{\tiny \mbox{FS}} - \frac{\hbar}{2(1 + \bar{z}_k z^k)}\left(z^k \, \d\bar{z}_k + \bar{z}_k \d z^k \right)	\otimes \frac{\partial}{\partial S} 
	\eeq
where $J_{\tiny \mbox{FS}}$ is given in~\eqref{Com-st-FS}.
It is not hard then to prove that $(\eta, g, \varphi)$ provides a \emph{Sasakian structure}  on  the extended space $\CP(\h_0) \times \R$~\cite{sasaki1960, sasaki1960-II,boyer2008sasakian,blair2010riemannian}, which is the analogue of the K\"ahler structure on $\CP(\h_0)$. 


\subsection{Dynamics: contact Hamiltonian systems on $\CP(\h_0)\times\mathbb R$}\label{subsec:CHDCPh}

The contact form allows one
to associate with every smooth function $\H\in \F(\CP(\h_{0}) \times \R)$ a Hamiltonian vector field $X_{\tiny \H}$, defined by 
	\beq\label{Diss-Lag-G}
	- i_{X_{\tiny \H}} \d \eta = \d \H - (\pounds_\xi \H) \eta
	\qquad
	\text{and}
	\qquad
	i_{X_{\tiny \H}} \eta = - \H \, .
	\eeq
In this case $\H$ is called the \emph{contact Hamiltonian function}. 
Now we proceed to compute the coordinate expression of $X_{\tiny \H}$, which in general may be written as 
	\beq
	X _{\tiny \H} = X_{z^k} \frac{\partial }{\partial z^k} 
	+ X_{\bar{z}_k} \frac{\partial }{\partial \bar{z}_k} + X_S \frac{\partial }{\partial S} \, ,
	\eeq
where the component $X_{\bar{z}_k}$ is the complex conjugate of $X_{z^k}$. 
Using~\eqref{one-form} and~\eqref{Diss-Lag-G} one finds that the components of the contact Hamiltonian vector field 
in these coordinates
are
	\begin{align}
	X_{z^k} & =  - \frac{\i}{\hbar} \, (1 +  |{\bf z}|^2 ) 
	\left( 
	\frac{\partial \H}{\partial \bar{z}_k} + z^k \, \bar{z}_l \, \frac{\partial \H}{\partial \bar{z}_l} 
	 \right) 
	 + \frac{z^k}{2}(1 +  |{\bf z}|^2 ) \frac{\partial \H}{\partial S} \, , \\
	X_S & = - \H - \frac{1}{2} (1 +   |{\bf z}|^2 )
	\left(
	\bar{z}_k \frac{\partial \H}{\partial \bar{z}_k} + z^k \frac{\partial \H}{\partial z^k} 
	\right) \, ,
	\end{align}
which implies that the integral curves are solutions to the system of the differential equations
	\begin{align} \label{Contact-Eq-Mot}
	\dot{z}^k & = - \frac{\i}{\hbar} \, (1 +   |{\bf z}|^2 ) 
	\left( 
	\frac{\partial \H}{\partial \bar{z}_k} + z^k \, \bar{z}_l \, \frac{\partial \H}{\partial \bar{z}_l} 
	 \right) 
	 + \frac{z^k}{2}(1 +   |{\bf z}|^2 ) \frac{\partial \H}{\partial S} \, , \nonumber  \\
	\dot{\bar z}_k & = \frac{\i}{\hbar} \, (1 +   |{\bf z}|^2 ) 
	\left( 
	\frac{\partial \H}{\partial z_k} + \bar{z}_k \, z^l \, \frac{\partial \H}{\partial z^l} 
	 \right) 
	 + \frac{\bar{z}_k}{2}(1 +  |{\bf z}|^2 ) \frac{\partial \H}{\partial S} \, , \nonumber \\
	\dot{S} & = - \H - \frac{1}{2} (1 +  |{\bf z}|^2 )
	\left(
	\bar{z}_k \frac{\partial \H}{\partial \bar{z}_k} + z^k \frac{\partial \H}{\partial z^k} 
	\right) \, .
	\end{align}
From these, we can compute the evolution of any arbitrary real function $\mathscr{F} \in \F(\CP(\h_{0}) \times \R)$, to obtain
	\begin{equation}\label{eq:evolutionofF}
	\frac{\d \mathscr{F}}{\d t}  = X _{\tiny \H}[\mathscr{F}] 
	 =  -\H \pounds_{\xi} \mathscr{F} + \Lambda(\d \mathscr{F},  \d \H ) \, ,
	\end{equation}
where the bivector $\Lambda$ is given by
	\beq
	\Lambda = (1 +   |{\bf z}|^2 )  \Bigg[
	- \frac{\i}{\hbar}
	\left(
	\frac{\partial  }{\partial z^k} \wedge \frac{\partial }{\partial \bar{z}_k} 
	+
	z^k \, \frac{\partial }{\partial z^k} \wedge \bar{z}_l \frac{\partial }{\partial \bar{z}_l}  
	\right)
	+ \frac{1}{2} \,
	\frac{\partial}{\partial S} \wedge \left( z^k \frac{\partial}{\partial z^k} + \bar{z}_k \frac{\partial }{\partial \bar{z}_k} 
	\right)
	\Bigg] \, .
	\eeq
Because in this work we are interested in characterizing dissipative systems that dissipate the energy 
of some declared reference Hamiltonian system, we shall henceforth assume that the contact Hamiltonian $\H$ may be written as
	\beq \label{Cont-En}
	\H = e_{\mathbf{H}} + f(S)\,,
	\eeq
where the first term $ e_\mathbf{H}$ is the expectation value of the Hamiltonian operator $\mathbf{H}$ of the conservative reference system, 
given in Eq.~\eqref{Exp-Val-Ham}, 
while the second term $f(S)$ is a perturbation on the system giving an effective characterization of the interaction between the conservative system and the environment.
Therefore, considering the expectation value $e_\mathbf{A} \in \F(\CP(\h_{0}))$ of an arbitrary observable $\mathbf{A}$ of the reference system, from~\eqref{eq:evolutionofF}
we obtain
	\beq \label{Cont-Time-Ev}
	\frac{\d e_{\mathbf{A}}}{\d t} = \{ e_\mathbf{A} , e_\mathbf{H}  \}_{\omega_{\tiny \mbox{FS}}} +
	\frac{1}{2} (1 +  |{\bf z}|^2 ) 
	\left( z^k \frac{\partial e_\mathbf{A}}{\partial z^k} 
	+ \bar{z}_k \frac{\partial e_\mathbf{A}}{\partial \bar{z}_k} \right) 
	f'(S)  \, ,
	\eeq
where $\{ \, \cdot \, , \, \cdot \, \}_{\omega_{\tiny \mbox{FS}}}$ is the Poisson bracket given in Eq.~\eqref{Exp-Pois-Bra}.
Hence, it should be clear that the expectation value of the energy $e_\mathbf{H}$ is not necessarily preserved along the trajectories of the contact Hamiltonian system.

Finally we remark that, from Eqs.~\eqref{Contact-Eq-Mot},
the contact dynamics $X_{\tiny \H} \in \X( \CP(\h_{0}) \times \R)$ is projectable onto a vector field $X \in \X(\CP(\h_{0}))$ 
if and only if the contact Hamiltonian function $\H$ 
is linear in $S$.  This is because only in this case the equations for the variables $z^{k}$ that characterize the state of the reference system 
defined on $\CP(\h_{0})$ are
decoupled from the equation for the additional variable $S$.
It is for this reason that in the following we will focus on this case only, leaving a detailed analysis of the general case to future investigations.

Now, in analogy with the symplectic case,
we can use the Riemannian structure introduced in~\eqref{eq:gFSE}
to compute gradient vector fields and the analogue of the Jordan product in the extended space.

As usual, the gradient vector field of a function $ \mathscr{A} \in\F( \CP(\h_{0}) \times \R)$ is defined by
	\beq\label{eq:GVFE}
	 g(Y_{\mathscr{A}}, \, \cdot \, )=\d \mathscr{A}\,.
	\eeq
In order to obtain a relationship similar to~\eqref{eq:relJ}, we can use the corresponding Hamiltonian vector
field $X_{\mathscr{A}}$ and the relationship
	\beq \label{Prop-1} 
	- \d \eta (X_{\mathscr{A}} , \, \cdot \,)= g (\varphi X_{\mathscr{A}} , \, \cdot \,) \, 
	\eeq
into the left hand side of Eq.~\eqref{Diss-Lag-G} to obtain
	\beq \label{def-grad}
	g (\varphi X_{\mathscr{A}} , \, \cdot \, ) = \d \mathscr{A} - (\pounds_\xi \mathscr{A}) \eta\,.
	\eeq
Finally, because $g(\xi , \, \cdot \,) =~\frac{2}{\hbar}\eta$, we conclude that
	\beq
	Y_{\mathscr{A}} =  \varphi X_{\mathscr{A}} + \hbar \, (\pounds_\xi \mathscr{A}) \, \xi \, .
	\eeq
In local coordinates we obtain
	\beq
	Y_{\mathscr{A}} = Y_{z^k} \, \frac{\partial}{\partial z^k}
	+ Y_{\bar{z}_k} \, \frac{\partial}{\partial \bar{z}_k} 
	+ Y_S \frac{\partial}{\partial S} \, ,
	\eeq 
where the component $Y_{\bar{z}_k}$ is the complex conjugated of $Y_{z^k}$ and
	\begin{align}
	Y_{z^k} & =  - \frac{1}{\hbar} \, ( 1 + \bar{z}_k \, z^k) 
	\left( 
	\frac{\partial \mathscr{A}}{\partial \bar{z}_k} + z^k \, \bar{z}_l \, \frac{\partial \mathscr{A}}{\partial \bar{z}_l} 
	 \right) 
	 + \frac{z^k}{2 \i}(1 + \bar{z}_k \, z^k) \frac{\partial \mathscr{A}}{\partial S} \\
	Y_S & = \frac{1}{2 \i} \, ( 1 + \bar{z}_k \, z^k )
	\left(
	z^k \frac{\partial \mathscr{A}}{\partial z^k} - \bar{z}_k \frac{\partial \mathscr{A}}{\partial \bar{z}_k}  
	\right)
	+ \frac{\hbar}{2} ( 1 -  \, \bar{z}_k \, z^k) \frac{\partial \mathscr{A}}{\partial S} \, .
	\end{align}
Now, the analogue of the Jordan bracket~\eqref{eq:Jordan} is the \emph{extended Jordan product}
	\beq\label{eq:JordanE}
	\{\mathscr{A},\mathscr{B}\}_{g}=g(Y_{\mathscr{A}},Y_{\mathscr{B}})\,,
	\eeq
which in coordinates reads
	\beq \label{Grad-Deriv}
	\{\mathscr{A},\mathscr{B}\}_{g}= G( \d \mathscr{A} , \d \mathscr{B} ) 
	+ \frac{\hbar}{2} ( 1 -  \, \bar{z}_k \, z^k) \frac{\partial \mathscr{A}}{\partial S} \frac{\partial \mathscr{B}}{\partial 	S} \, ,
	\eeq	
where the symmetric bivector $G$ is defined as 
	\beq\label{eq:G}
	G = ( 1 + \bar{z}_k \, z^k )  \Bigg[
	- \frac{1}{\hbar}
	\left(
	\frac{\partial  }{\partial z^k} \otimes_{\tiny \mbox{S}} \frac{\partial }{\partial \bar{z}_k} 
	+
	z^k \, \frac{\partial }{\partial z^k} \otimes_{\tiny \mbox{S}} \bar{z}_l \frac{\partial }{\partial \bar{z}_l}  
	\right)
	+
	\frac{1}{2 \i} \,
	\frac{\partial}{\partial S} \otimes_{\tiny \mbox{S}} \left( z^k \frac{\partial}{\partial z^k} - 
	\bar{z}_k \frac{\partial }{\partial \bar{z}_k} 
	\right)
	\Bigg] \, .
	\eeq
Thus we can rewrite Eq.~\eqref{Grad-Deriv} in a more explicit form as
	\begin{align}\label{eq:JordanEexpr}
	 \{\mathscr{A},\mathscr{B}\}_{g}& =  \{ \mathscr{A} , \mathscr{B} \}_{g_{\tiny \mbox{FS}}} 
	+ \frac{1}{2 \i} ( 1 + \bar{z}_l \, z^l ) \left[
	\frac{\partial \mathscr{A} }{\partial S}
	\left( z^k \frac{\partial \mathscr{B}}{\partial z^k} - \bar{z}_k \frac{\partial \mathscr{B}}{\partial \bar{z}_k} \right) 
	+ \frac{\partial \mathscr{B} }{\partial S}
	\left( z^k \frac{\partial \mathscr{A} }{\partial z^k} - \bar{z}_k \frac{\partial \mathscr{A} }{\partial \bar{z}_k} \right) 
	\right] 
	\nonumber \\ &
	+ \frac{\hbar}{2} ( 1 -  \, \bar{z}_k \, z^k) \frac{\partial \mathscr{A}}{\partial S} \frac{\partial \mathscr{B} }{\partial S} \, ,
	\end{align}

Finally, we can generalize the relationship between the uncertainties and the Jordan product, Eqs.~\eqref{varianza} and~\eqref{correlation}, by \emph{defining}
	\beq \label{contact-uncertainty}
	\sigma^2_\mathscr{A} = - \frac{\hbar}{2}\{\mathscr{A},\mathscr{A}\}_{g}
	\eeq
and
	\beq\label{contact-uncertainty2}
	\sigma_{\mathscr{A} \mathscr{B}} = - \frac{\hbar}{2}\{\mathscr{A},\mathscr{B}\}_{g}\,.
	\eeq
Here we should remark that $\sigma^2_\mathscr{A}$ and $\sigma_{\mathscr{A} \mathscr{B}}$ in general 
(when either $\frac{\partial\mathscr{A}}{\partial S}\neq 0$ or $\frac{\partial\mathscr{B}}{\partial S}\neq 0$)
do not seem to have any immediate statistical meaning. 
However, when both
$\mathscr{A}$ and $\mathscr{B}$ are basic functions (i.e.~when $\frac{\partial\mathscr{A}}{\partial S}=\frac{\partial\mathscr{B}}{\partial S}=0$)
they are exactly the variance and the correlation of the corresponding operators, as it can be seen by 
using~\eqref{eq:JordanEexpr} and comparing with~\eqref{varianza} and~\eqref{correlation}.
Interestingly, even in such case the evolution is different in general, 
as $z(t)$ in~\eqref{contact-uncertainty} and~\eqref{contact-uncertainty2}
evolves according to the (dissipative) contact Hamiltonian equations of motion.
	

\subsubsection*{Example: the dissipative qubit}
 
As an example, let us introduce dissipation in the qubit system described in the previous section. 
To do so, we consider the contact Hamiltonian
 	\beq\label{Cont-Ham-Lin-S} 
	\H = e_\mathbf{H} - \gamma S \, ,
	\eeq
where $e_\mathbf{H}$ is the expectation value of the Hamiltonian of the conservative qubit, Eq.~\eqref{exp-value}, 
and $\gamma$ is a positive real constant which quantifies the strength of the coupling between the conservative system and the environment.

The Hamiltonian equations of motion~\eqref{Contact-Eq-Mot} associated with the contact Hamiltonian \eqref{Cont-Ham-Lin-S} read
	\begin{align}
	\dot{z} & = \frac{\i}{\hbar} \left[ \bar{V} z^2 - (H_1 - H_2) z - V \right] - \frac{\gamma}{2} z (1 +  |{z}|^2) 
	\label{z-eq}\\
	\dot{S} & = - \H - \frac{1}{2} \, \left[ \frac{1 -  |{z}|^2}{1 + |{z}|^2} \right] (\bar{V} z + V \bar{z}) 
	+ \frac{ |{z}|^2}{1 + |{z}|^2} (H_2 -H_1)		\, .
	\end{align}
We see that, as expected,
Eq.~\eqref{z-eq} provides an {effective} equation of motion for the variable $z$, describing the state of the reference system, 
which is {decoupled} from the equation for the additional variable $S$.

As in the conservative case, we may now study the qualitative behavior of the dissipative qubit, 
focussing only on Eq.~\eqref{z-eq}.
For simplicity, we consider here as an example the particular case $H_1 = H_2$. 
The critical points $z_s$ satisfy
	\beq \label{SIng-Eq}
	\frac{\i}{\hbar} \left[ \bar{V} z_s^2 - V \right] - \frac{\gamma}{2} z_s (1 + |z_s|^2) = 0 	\, .
	\eeq
To find solutions to this algebraic equation, first we assume that $|z_{s}|=1$, to obtain the reduced second-order equation
	\beq
	 \overline{V} z_s^2 + \i \, \hbar \, \gamma z_s - V = 0,
	\eeq
with solutions
	\beq \label{Diss-Sing-Point}
	z^{(1,2)}_s = \frac{ - i \hbar \gamma 
	\pm \sqrt{\Delta_{-} } }{2 \, \overline{V}},
	\quad
	\eeq
where $\Delta_{-} := 4 |V|^2 - \hbar^2 \gamma^2\geq 0$.
A direct computation shows that $|z^{(j)}_s |= 1$, for $j = 1,2$ if and only if $\Delta_{-}\geq 0$, and therefore these two critical points exist only in such case.
When they exist, in the conservative limit $\gamma \to 0$ one recovers the analogous solutions in Eq.~\eqref{Cons-Sing-Point} for the case $H_1 = H_2$.
Additionally, in the dissipative qubit, one finds the critical point  
	\beq
	z^{(3)}_s = - \frac{2 \i V}{\hbar \gamma} \,,
	\eeq
which always exists. 

To continue the qualitative study of the dynamics, 
we notice that the eigenvalues of the linearized system at the critical points $z^{(1)}_s$ and $z^{(2)}_s$ are always the same and they are given by
	\beq
	\lambda_{\pm}^{(1,2)} = - \frac{\gamma}{2} \pm \i \sqrt{ \frac{\Delta_{-}}{\hbar^2} - \frac{\gamma^2}{4}},
	\eeq
while at the critical point $z^{(3)}_{s}$ we have the eigenvalues
	\beq
	\lambda_{\pm}^{(3)} = - \frac{\hbar^{2}\gamma^{2}\pm 4|V|^2}{2\hbar^{2}\gamma}\,,
	\eeq
which are always real.
From this linearization it is possible to conclude that there is a bifurcation depending on the value of $\gamma$:

	\begin{itemize}
	\item[i)] for $\gamma$ such that $\Delta_{-}<0$ we have only the critical point $z^{(3)}_{s}$, which is a stable 	node;
	\item[ii)] for $\gamma$ such that $\Delta_{-}=0$ we have two critical points $z^{(1)}_{s}=z^{(2)}_{s}$ and 		$z^{(3)}_{s}$, 
	and both are non-hyperbolic (eigenvalues $\{0,\textcolor{blue}{-}\gamma\}$);
	\item[iii)] for $\gamma$ such that $\Delta_{-} > 0$ we have three different critical points with the following 		behavior: $z^{(3)}_{s}$ is always a saddle; 
	the behavior of $z^{(1)}_{s}$ and  $z^{(2)}_{s}$ depends on the term $\frac{\Delta_{-}}{\hbar^{2}}-			\frac{\gamma^{2}}{4}$. If this is positive, then the critical points are stable foci, while if this term is non-positive 	then they are stable nodes.
	\end{itemize}

Note that even in this simple example we get a very interesting dynamical behavior. Moreover, we remark that
these types of critical points cannot be obtained in the case of a conservative (unitary) evolution.

\begin{figure}[h!]
	\centering
	\includegraphics[width = 7. cm]{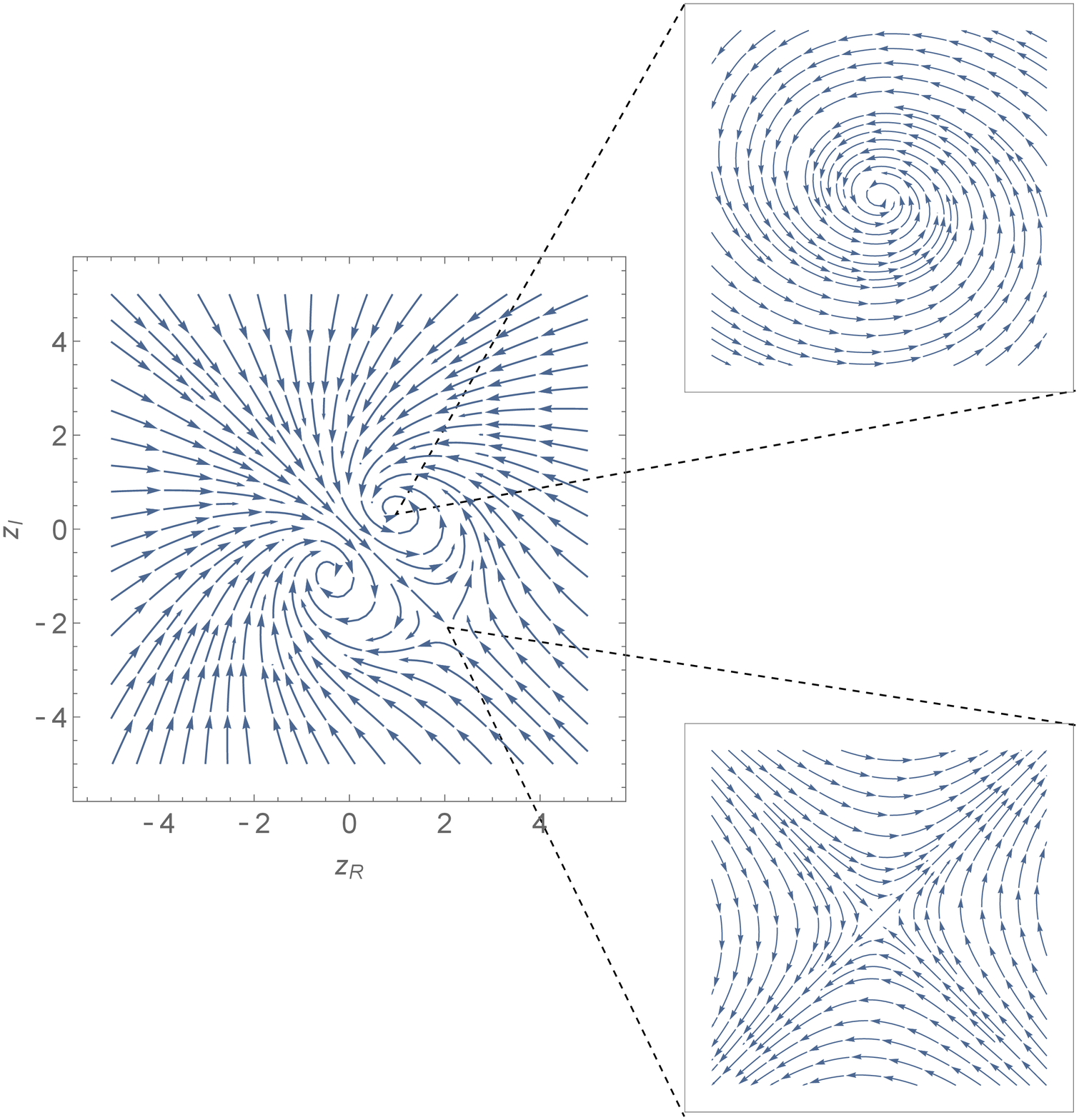}
	\qquad\qquad
	\includegraphics[width = 6.5 cm]{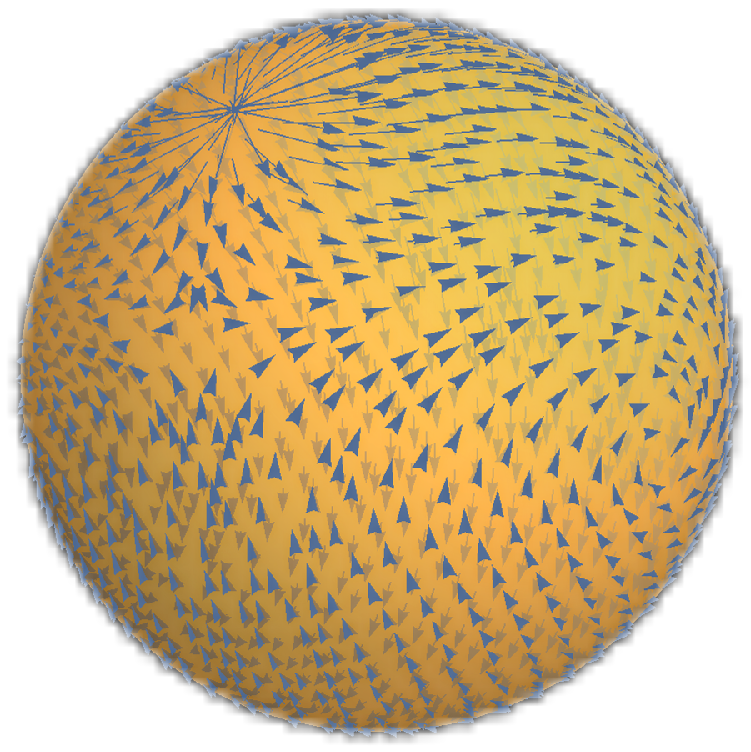}
	\caption{Phase portrait of the dissipative qubit with $H_1 = H_2$ both in the complex plane and on $S^{2}$. The parameters chosen for this representation are the following: $V = 1 + \i$ and $\gamma = 1$. One may see the two stable foci and the saddle point both in the plane and in $S^{2}$, together with the unstable node at the north pole of the sphere, which cannot be seen in this chart on the plane. More details in the main text.}
	\label{Fig-3}
	\end{figure}

As a way of example, we have depicted in Fig.~\ref{Fig-3} the phase portrait for a choice of $\gamma$ such that $\Delta_{-}>0$ and $\frac{\Delta_{-}}{\hbar^{2}}-\frac{\gamma^{2}}{4}>0$.
As can be seen in first panel in Fig.~\ref{Fig-3}, we have two stable foci and a saddle.
However, recalling that this chart covers only the part of the Bloch sphere $S^{2}$ excluding the north pole, 
in the second panel in Fig.~\ref{Fig-3} we also depict the corresponding vector field on $S^{2}$,
from which it is clear that there is an additional critical point at the north pole, which is an unstable node.

Finally, taking into account~\eqref{contact-uncertainty} we may compute the uncertainty of the Hamiltonian operator, given by
	\beq\label{Dissipative-Unc}
	\sigma^2_{\mathbf{H}\text{q}} =  \frac{1}{(1 + | {z} |^2)^2} 
	| V \bar{z}^2 - (H_1 - H_2) \bar{z} - \overline{V} |^2 \,.
	\eeq
We stress that, although the functional forms of~\eqref{Conservative-Unc} and~\eqref{Dissipative-Unc} are the same, 
they are different functions of time, since in the two cases $z(t)$ evolves according to different dynamical equations.	
	

\section{The contact master equation}\label{sec:CHDM}
 
So far we have considered dissipative quantum evolutions by using contact Hamiltonian systems defined on the  manifold $\CP(\h_{0})\times\R$.
However, the most common description of dissipative quantum systems is by means of density operators, and therefore in this section
we shall adapt our approach to this setting.

Given a $C^\ast$-algebra $\A$, \emph{a state} $\rho$ on $\A$ is a continuous linear function in the dual $\A^\ast$ of $\A$ 
such that for all observables $\mathbf{a}  \in \A$ one has that
	\beq
	\rho(\mathbf{a}) \in \R\, ,
	\quad
	\rho(\mathbf{a} \mathbf{a}^\dagger) \geq 0 \, ,
	\quad
	\text{and}
	\quad
	\rho(\mathbb{I}) = 1 \, ,
	\eeq
where $\mathbb{I} \in \A$ is the identity and an \emph{observable} $\mathbf{a}\in \A$ is a self-adjoint element.
In particular, for finite-dimensional systems, we may restrict our attention to the $C^\ast$-algebra of $n \times n$ complex matrices, 
i.e.~we may consider $\a_n = M_n(\C)$ with $n \geq 2$. 
Thus, the space of states $\mathcal{S}$ of $\a_n$ is identified as
	\beq
	\mathcal{S} := \left\{ \rho \in \D_n^\ast \subset \a^\ast_n \, | \, \rho(\mathbf{a} \mathbf{a}^\dagger) \geq 0 \, , 
	 \text{for all} \, \mathbf{a} \in \D_n \, ,
	\rho(\mathbb{I}) = 1 \, \right\} \, ,
	\eeq
where $\D_n \subset \a_n$ denotes the space of observables. 
The \emph{pairing map} $\mu:\D_{n} \times \mathcal{S} \to \R$ is the evaluation of the state $\rho$ on the self-adjoint element $\mathbf{a}$, and is given explicitly by
	\beq
	\mu(\mathbf{a} , \rho) \mapsto \rho(\mathbf{a}) := \Tr\{ \rho \, \mathbf{a} \} \, ,
	\eeq
which corresponds to the mean value for the outcome of the measurement of the observable $\mathbf{a}$ when the system is in the state $\rho$.
Furthermore, because there is a one-to-one correspondence between elements in $\D_n^\ast$ and $\D_n$, 
then it follows that the space $\mathcal{S}$ may be decomposed as
	\beq
	\mathcal{S} = \bigsqcup_{k = 1}^n \mathcal{S}_k \, ,
	\eeq
where $\mathcal S_k=\{\rho \in \mathcal{S}|\, \text{rk}(\rho) = k\}$. 
It is proven in Refs.~\cite{ciaglia2017dynamical, chruscinski2019stratified} that  every $\mathcal{S}_k$ is a \emph{homogeneous space} for the Lie group $SL( \a_n )$ and 
thus every $\mathcal S_k$ admits the structure of a differential manifold. Indeed, they are K\"ahler manifolds (see Refs.~\cite{ciaglia2017dynamical, chruscinski2019stratified} for details).
	
Here we are only interested in pure states, i.e.~rank-one projectors in $\mathcal{S}_1$.
Then, as we have already mentioned in Section~\ref{sec:GQM}, 
there is a one-to-one correspondence between states $|\psi] \in \CP(\h_{0})$ 
and rank-one projectors $\rho \in \mathcal{S}_1$, given by the relation \eqref{st-mat}. 
Thus in homogeneous coordinates we have
	\beq \label{z-rep}
	 | \psi ] =\frac{1}{\sqrt{1 + |{\bf z}|^2 }}\left(
	\begin{array}{c}
	  {\bf z}  \\
	  1
	\end{array}
	\right)\, ,
	\quad
	\text{then}
	\quad
	\rho_\psi = | \psi ] [ \psi | 
	= \frac{1}{1 + |{\bf z}|^2 }
	\left(
	\begin{array}{ccc}
	{\bf z} {\bf z}^\dagger   & {\bf z} \\
	{\bf z}^\dagger    & 1 
	\end{array}
	\right) \, .
	\eeq 
Now, starting from the expression for the density matrix in the second equation in Eq.~\eqref{z-rep} 
and taking into account the equations of motion \eqref{Contact-Eq-Mot}, one may deduce the equation of motion for $\rho$,
which takes the general form
	\beq\label{eq:dotrho1}
	\dot{\rho} = \frac{\i}{\hbar}
	\left(
	\begin{array}{cc}
	z^j \frac{\partial \H }{\partial z^k} - \bar{z}_k \frac{\partial \H }{\partial \bar{z}_j}
	& - \frac{\partial \H }{\partial \bar{z}^j} - z^j z^l \frac{\partial \H }{\partial z^l} \\
	&  \\
	\frac{\partial \H }{\partial z^k} + \bar{z}_k \bar{z}_l  \frac{\partial \H }{\partial \bar{z}_l}
	&-  z^l \frac{\partial \H }{\partial z^l} + \bar{z}_l \frac{\partial \H }{\partial \bar{z}_l}
	\end{array}
	\right)
	+ \frac{\partial \H}{\partial S} 
	\left(
	\begin{array}{cc}
	\frac{z^j \, \bar{z}_k }{1 + |{\bf z}|^2 } 
	& \frac{z^j}{2} \, \frac{1 - |{\bf z}|^2}{1 + |{\bf z}|^2 } \\
	&  \\
	\frac{\bar{z}_k}{2} \, \frac{1 - |{\bf z}|^2}{1 + |{\bf z}|^2 }
	& -\frac{|{\bf z}|^2}{1 + |{\bf z}|^2 } 
	  \end{array}
	\right) \, .
	\eeq
	
We first observe from the above equation that $\Tr\{ \, \dot{\rho} \, \} = 0$ independently of the choice of the contact Hamiltonian.
Therefore, the trace of $\rho$ is preserved along the evolution, which is a fundamental statistical condition for any admissible quantum evolution.

Remarkably, 
the evolution~\eqref{eq:dotrho1} can be written using two brackets, similar to the structure of the metriplectic and  GENERIC formalisms~\cite{morrison1984bracket,guha2007metriplectic,morrison2009thoughts,grmela1997dynamics,pavelka2018multiscale} 
(although an approach to dissipative quantum systems directly based on the analogy with the GENERIC equation leads to a different
type of master equation, see~\cite{ottinger2011geometry,ottinger2010nonlinear}).
Indeed, we have (see Appendix~\ref{appA} for the proof)
	\beq \label{Rho-Ev}
	\dot{\rho} = \frac{\i}{\hbar} [ \, \rho \, , \, \mathbf{H} \, ]
	+  \frac{\partial \H}{\partial S} \, \left[ \rho, {\mathbf A}({\bf z}) \right]_+ \,,
	\eeq 
where the dissipative bracket $[ \cdot \, , \, \cdot ]_{+}$ is the anti-commutator and we have defined the \emph{dissipative potential} to be the Hermitian operator $\mathbf{A}({\bf z})$ 
given by
	\beq
	\mathbf{A}({\bf z}) = \left(
	\begin{array}{cc}
	\mathbb{A} & \mathbf{w}({\bf z}) \\
	&  \\
	\mathbf{w}^\dagger({\bf z}) & A({\bf z})
	  \end{array}
	\right) \,,
	\eeq
\noindent where $\mathbb{A}$ is an arbitrary $(n-1) \times (n-1)$-dimensional Hermitian matrix, $\mathbf{w}$ is the $(n-1)$-dimensional column vector 
given by
	\begin{equation}
	\mathbf{w}({\bf z}) = \frac{1}{1 + |{\bf z}|^2} \left[ \left(1 + |{\bf z}|^2\right) \mathbb{I} - {\bf z} {\bf z}^\dagger  \right]
	\left[ \frac{1}{2}\left( 1 + |{\bf z}|^2\right)\mathbb{I}  - \left( {\bf z}^\dagger \, \mathbb{A} \,  {\bf z} \right) \mathbb{I} - \mathbb{A} \right] {\bf z}.
	\end{equation} 
and $A({\bf z})$ is the real quantity 
	\begin{equation}
	A({\bf z})= {\bf z}^\dagger \, \mathbb{A} \, {\bf z} - |{\bf z}|^2 \, .  
	\end{equation} 	

Let us remark that in~\eqref{Rho-Ev} there is a conservative part given by the standard von Neumann equation, plus a contact perturbation introducing dissipation 
in the  conservative reference dynamics. 
Moreover, in general equation \eqref{Rho-Ev} is coupled to the equation of motion for the variable $S$ given by 
	\beq \label{Contact-S}
	\dot{S} = - \H - \frac{1}{2} (1 + |{\bf z}|^2)
	\left(
	\bar{z}_k \frac{\partial \H}{\partial \bar{z}_k} + z^k \frac{\partial \H}{\partial z^k} 
	\right) \,,
	\eeq
and  they decouple if and only if the contact Hamiltonian $\H$ is linear in $S$.
In such case one obtains from \eqref{Rho-Ev} a Markovian equation for $\rho$, which we call the \emph{contact master equation}.
This is the case we will address in the following, leaving a detailed study of the general case to future works.



Let us now compare the contact master equation~\eqref{Rho-Ev} with the GKLS equation normally employed to describe dissipative phenomena in quantum systems.
For the GKLS equation we have
	\beq \label{GKLS-Ev}
	\dot{\rho} =  \frac{\i}{\hbar} [ \, \rho \, , \, \mathbf{H} \, ]
	- \frac{1}{2} \sum_j
	\left[ \rho \, , \, \mathbf{V}_j^\dagger \mathbf{V}_j \right]_{+}
	+  \sum_j  \mathbf{V}_j \rho \mathbf{V}_j^\dagger \, ,
	\eeq
where $\mathbf{V}_j$ and $\sum_j  \mathbf{V}_j \mathbf{V}_j^\dagger$ are bounded operators.
First of all, we  note that by comparing the GKLS evolution in~\eqref{GKLS-Ev} with the contact evolution in \eqref{Rho-Ev}, 
it is clear the absence in the latter of the \emph{jump term}
$ \sum_j  \mathbf{V}_j \rho \mathbf{V}_j^\dagger$ (also known as the \emph{Choi-Kraus term}~\cite{chruscinski2017brief}).
This is expected because this term is the one responsible for the change of the rank of the density matrix, and the contact evolution considered here preserves the purity of the states.
However, while in the GKSL equation the jump term is needed in order to enforce conservation of the trace and the complete-positivity of the map, 
in the contact evolution we see that it is no longer needed, for these two properties are both automatically satisfied (at least for pure states).
Hence we see that the contact master equation provides a (nonlinear) description of coherent dissipative phenomena that satisfies all the statistical requirements of quantum mechanics.

Furthermore, one may also compare the contact master equation with other descriptions normally employed to describe dissipative systems. 
For instance, one may consider the description of an \emph{optical MASER} (Microwave Amplification by Stimulated Emission of Radiation)~\cite{lamb1964, carmichael2009}.
Considering for simplicity the case of a 2-state system, one introduces the non-normalized density matrix
	\beq 
	\rho =
	\left(
	\begin{array}{c c}
	\psi^1 \bar{\psi}^1 &  \psi^1 \bar{\psi}^2\\
	& \\
	\psi^2 \bar{\psi}^1 & \psi^2 \bar{\psi}^2
	\end{array}
	\right) \,,
	\eeq
whose equation of motion reads 	
	\beq \label{Rho-SL}
	\dot{\rho} =  \frac{\i}{\hbar} [ \, \rho \, , \, \mathbf{H} \, ] 
	- \frac{1}{2}
	\left[ \rho \, , \, \mathbf{\Gamma} \right]_{+} \,,
	\eeq
where $\mathbf H$ is a time-dependent Hermitian operator and $\mathbf{\Gamma}$ is a diagonal operator with entries $\gamma_{1},\gamma_{2}>0$ that 
describes phenomenologically the radiative decays of the eigenstates of  $\mathbf H$ to the ground state.
We observe that the positivity of $\mathbf{\Gamma}$ implies that $\Tr\{ \dot{\rho} \} < 0$ and therefore the evolution is not trace-preserving. 
It is also interesting to observe that the dynamics \eqref{Rho-SL} may be rewritten via the relation \eqref{z-rep} as a differential 
equation on $\CP(\h_{0})$, which reads
	\beq
	\dot{z} = \frac{\i}{\hbar} \left[ \overline{V} z^2 - (H_1 - H_2) z - V \right] 
	+ \frac{1}{2 \hbar} (\gamma_1 -  \gamma_2 ) \, z \,.
	\eeq
This equation should be compared with the contact one in Eq.~\eqref{z-eq}. Indeed, by considering $\gamma =  \frac{1}{\hbar} (\gamma_1 -  \gamma_2 )$
we see that one has the same equation, up to the nonlinear term $\frac{\gamma}{2} z^2 \bar{z}$.
We conclude that the pathologies of Eq.~\eqref{Rho-SL} can be removed in at least two ways:
on the one side one can introduce the jump term $\mathbf{\Gamma} \, \rho \, \mathbf{\Gamma}$ 
to obtain an equation of GKLS type, but this leads to introducing dissipation of the rank, i.e.~introducing decoherence in the system;
on the other side, the contact master equation provides an alternative option, one that fixes the pathologies of Eq.~\eqref{Rho-SL} while preserving the coherence of the states. 
This however comes at the price of introducing nonlinearities.

\vsp

\section{Application: radiative decay}\label{sec:Example}

The simplest dissipative quantum phenomenon that one can describe by the contact evolution 
is the radiative decay of a 2-level atom, 
with levels $| 1 \rangle$ and $| 2 \rangle$. 
To show this, let us consider as the conservative system the Hamiltonian
	\beq \label{Exam}
	\mathbf{H} = \left(
	\begin{array}{c c}
	H_1  &  0   \\
	& \\
	0 &  H_2   \\   
	\end{array}
	\right)
	\eeq
where $H_1$ and $H_2$ are the energies of the states, with $H_1 > H_2$.
Then the decay of the particle may be modelled by the contact Hamiltonian
	\beq \label{Cont-Ham-Decay}
	\H = e_\mathbf{H} - \gamma \, S
	\eeq
where $e_\mathbf{H}$ represents the expectation value of the Hamiltonian \eqref{Exam} 
and the constant damping factor $\gamma > 0$ describes phenomenologically the radiative decay from the state $| 1 \rangle$ to $| 2 \rangle$.
The corresponding contact Hamiltonian equations of motion can be obtained from Eq.~\eqref{Contact-Eq-Mot} and read
	\begin{align}
	\dot{z} & = - \frac{\i}{\hbar} (H_1 - H_2) z - \frac{\gamma}{2} z (1 + |z|^2) \, , \\ 
	\dot{S} & = - \H - \frac{|z|^2}{1 + |z|^2} (H_1 - H_2)	 \label{S-decay}\, ,
	\end{align}
i.e.~we have a decoupled system of differential equations, as expected, and therefore we can focus on the first equation only, whose solution can be given explicitly by
	\beq \label{Sol-decay}
	z(t) = \frac{\e^{\i \varphi_0 }}{ \sqrt{ \e^{ \gamma t + 2 \kappa_0 } - 1}} 
	\e^{ - \frac{\i}{\hbar}(H_1 - H_2) t },
	\eeq
where the phase $\varphi_0$ and the real constant $\kappa_0$ are defined by the initial condition $z(0) = z_0$. 
Let us recall that because the homogeneous coordinates $z \in \CP(\h_{0})$ have been mapped to the sphere via the stereographic projection, then, in order to consider as the initial condition the state with energy $H_1$, one has to consider $z_0$ as the ``point at infinity'' in the plane, corresponding to the north pole of the sphere. In the representation~\eqref{Sol-decay} this corresponds to taking  $\kappa_0 \rightarrow  0$.

\begin{figure}[h!]
	\centering
	\includegraphics[width = 9 cm]{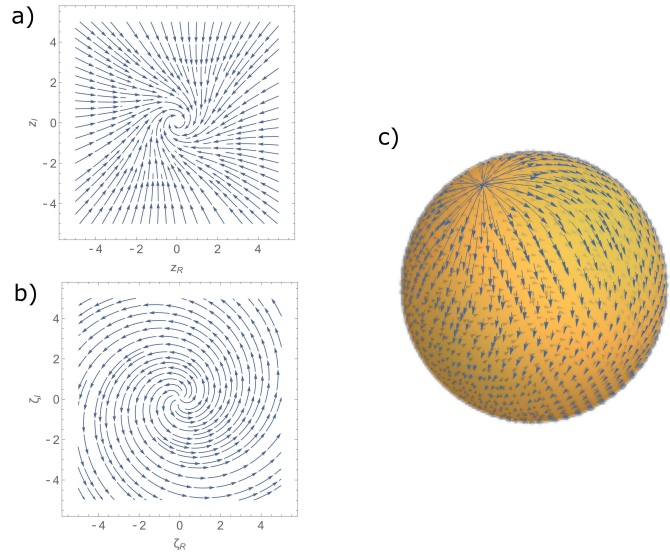}
	\caption{ Phase portrait of the vector field associated with  the quantum decay from the higher energy 
	$H_1 = 4$ to the lower energy $H_2 = 2$ with damping parameter $\gamma = 1$. }
	\label{Fig-5}
	\end{figure}

As an illustration, we depict the behaviour of the solutions of Eq.~\eqref{Sol-decay} in Fig.~\ref{Fig-5}. 
In Fig.~\ref{Fig-5}a we show the phase portrait in homogeneous coordinates in the plane
corresponding to the stereographic projection from the north pole of the sphere.
Here we have a stable focus at the origin and consequently all  nearby solutions evolve towards the origin.
On the other hand, considering the coordinates obtained by stereographic projection from the south pole, one has an unstable focus at the origin, Fig.~\ref{Fig-5}b. 
Finally, both charts form an atlas for the Bloch sphere $S^2$ and the behaviour of the vector field on such sphere is displayed in Fig.~\ref{Fig-5}c. 
All this is in agreement with our physical interpretation because the north pole corresponds to the excited state with energy $H_1$ and 
the south pole to the state with lower energy $H_2$, i.e.~we are dissipating energy until the system finally decays in the lower state.

\begin{figure}[h!]
	\centering
	\includegraphics[width = 14 cm]{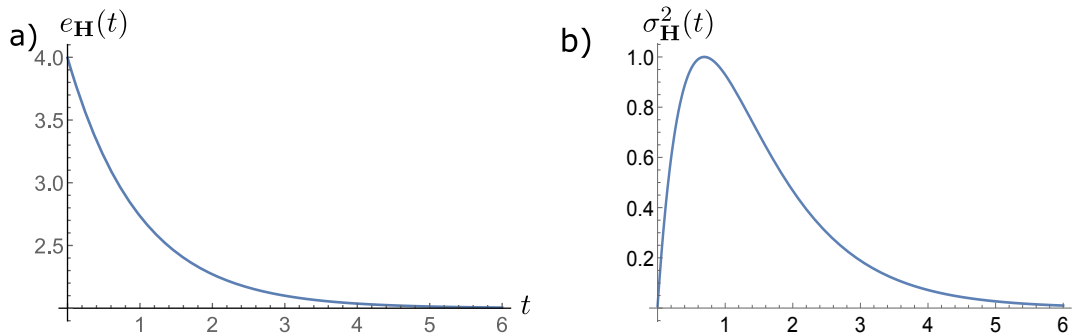}
	 \caption{a) Continuous transition from the state with energy $H_1 = 4$ to the state $H_2 = 2$ with damping 	parameter  $\gamma = 1$. 
	b) Time evolution of the uncertainty of the energy operator with the same parameters.}
	\label{Fig-6}
	\end{figure}

To observe clearly the continuous dissipation of the energy, 
one may compute the evolution of the expectation value of the conservative Hamiltonian operator \eqref{Exam}
using  the solution \eqref{Sol-decay}, to obtain
	\begin{equation}
	 e_\mathbf{H}(t) 
	  = \frac{1}{1 + | z(t)|^2 } \left[ \, 
	H_1 | z(t) |^2 + H_2 \right] 
	= H_2 + (H_1 - H_2) \e^{ -(\gamma t + 2 \kappa_0)} \,.
	\end{equation} 
Then it is not difficult to prove that for the initial condition $\kappa_0 \rightarrow 0$ we have $e_\mathbf{H}(t = 0)= H_1$ and
$	\lim_{t \to +\infty} e_\mathbf{H}(t) = H_2 \, ,$
thus describing the radiative decay from the excited to the ground state. 
In Fig.~\ref{Fig-6}a we display the continuous transition between these states, whose speed depends on the value of $\gamma$.
Furthermore, we remark that one may equivalently describe excitations within this model by simply exchanging $\gamma \to -\gamma$ 
and then having the transition 
from the lower state $H_2$ to the excited state $H_1$. 

One may also compute the uncertainty of the energy operator by means of the result in \eqref{contact-uncertainty}, namely
	\begin{equation}
	\sigma^2_\mathbf{H} 
	=  \frac{ |z(t)|^2}{(1 +  |z(t)|^2 )^2} (H_1 - H_2)^2 
	= \left[ 1 - \e^{ -(\gamma t + 2 \kappa_0)} \right] \e^{ - \gamma t - 2 \kappa_0} (H_1 - H_2)^2
	\, ,
	\end{equation}
whose evolution is plotted in Fig.~\ref{Fig-6}b. 
From this figure one may observe that the uncertainty of the energy starts at zero, then
increases up to the maximum value $\frac{1}{4}(H_1 - H_2)^2$ at $t = \frac{1}{\gamma} \ln{2}$, and after this maximum the curve decreases asymptotically to zero. 
Since $\sigma^2_\mathbf{H}$ is the statistical fluctuation around the expectation value, 
it follows that a measurement of $e_\mathbf{H}$ with complete certainty is only possible in principle at the initial and  final times.

Another relevant quantity is the probability of a transition between the states, denoted by $\mathscr{P}(t)$ 
and given by $\mathscr{P}(t) ~=~ {| \, [ \psi_0 | \psi_t ] \, |^2}$, where $| \psi_0 ]$ 
is the initial state and $| \psi_t ]$ corresponds to the evolved normalized state at time $t$.
Using~\eqref{z-rep} we obtain that $\mathscr{P}(t)$ is given by
	\beq
	\mathscr{P}(t) = \frac{|z(t) |^2}{1 + |z(t)|^2}\, ,
	\eeq
where in this expression we considered the initial condition $\kappa_0 \to 0$. 
Moreover, by means of Eq.~\eqref{Cont-Time-Ev} we can compute the rate of dissipation of this probability, which is
$\frac{\d \mathscr{P}}{\d t} = - \gamma \mathscr{P}$, meaning that one has an exponential decay $\mathscr{P}(t) = \e^{- \gamma t}$ for the probability of a transition. 

We remark again at this point that usually radiative decays are represented as \emph{sudden} ``jumps'' and not as continuous transitions.
However, recent experimental and theoretical works~\cite{minev2019catch,snizhko2020quantum} have shown 
that it is possible to see quantum jumps as continuous processes that preserve the coherence of the state. 
Here we have seen in this example that the contact master equation yields another possibility to model quantum decays (or excitations) as coherent and continuous processes.


\section{Conclusions and perspectives}\label{sec:Conclusions}

In this work we have put forward a novel approach to the description of  dissipative quantum systems,
based on the geometric approach to quantum mechanics and on the analogy with the description of 
classical dissipative systems based on contact Hamiltonian dynamics. 
Of special importance is the fact that the thus-obtained evolution dissipates the expectation value 
of the energy of the reference system, while preserving the purity of the states. In this way it yields a way 
to describe coherent dissipative dynamics, which inevitably escapes more standard approaches such
as the GKSL equation.
Among the possible applications, we have considered here in particular the important case of radiative decay
for a 2-level system, both because of its theoretical importance in understanding quantum mechanics and
because of recent experiments that point to an explanation in terms of the existence of coherent quantum 
trajectories for these systems~\cite{minev2019catch,snizhko2020quantum}.

At this point there is a number of interesting questions open for future work. 
In particular, in this effort we have considered the additional variable $S$ as a ``way to an end'', 
that is, as an effective tool that we have employed in order to produce a dissipative dynamics on the reference system in a geometric way. 
However, the term $\gamma S$ as we have used it has the dimensions of an energy and similar terms appear in the 
thermodynamic literature as the actual interchange of heat between the reference system and the environment~\cite{eberard2007extension,simoes2020contact}. 
Therefore, a deeper connection with an energy conservation principle may be responsible for our construction. 
Furthermore, the variable $S$ is linked to the action in contact systems by means of Herglotz' variational principle~\cite{georgieva2002first,georgieva2003generalized,liu2018contact,vermeeren2019contact,cannarsa2019herglotz}, and therefore it will be interesting to explore whether there is a connection with the action principles of quantum mechanics. 
Still concerning the variable $S$, in this work we have dealt only with contact Hamiltonian functions that are linear in $S$. This is because in this way we can directly guarantee that the resulting evolution on the space of pure quantum states respects all the principles of standard quantum mechanics. 
However, we have remarked that the evolution corresponding to the most general contact Hamiltonian still preserves the trace of the density operator.
This is interesting because in such case one expects to obtain a non-Markovian evolution which also agrees with the tenets of the quantum theory.

Another interesting problem is the extension of the present approach to infinite dimensional quantum systems. 
Solving this problem in all its generality is not an easy task; however, one may restrict to 
the problem of looking for the immersion of a contact manifold into the Hilbert space such that the evolution of the states is parametrized by the contact evolution.

Further still, given the importance of the Schr\"odinger dynamics in the description of quantum systems, 
one may wonder whether there is a Schr\"odinger-like equation associated with the contact evolution introduced here.
Indeed, for some particular cases it is possible to construct a Schr\"odinger equation; for instance, 
one may reproduce the contact evolution \eqref{z-eq} by means of the projection \eqref{Riccati-proj} starting from the nonlinear Schr\"odinger equations
	\beq
	\i \hbar
	\left(
	\begin{array}{c}
	  \dot{\psi}^1   \\
	  \\
	  \dot{\psi}^2  
	\end{array}
	\right)
	=
	\left(
	\begin{array}{ccc}
	H_1  &  V   \\
	& \\
	 \bar{V} &  H_2   \\   
	\end{array}
	\right)
	\left(
	\begin{array}{c}
	  \psi^1  \\
	  \\
	  \psi^2
	\end{array}
	\right) 
	+ \frac{\gamma}{2} 
	\left(
	\begin{array}{ccc}
	1  &  \frac{\psi^1}{\psi^2}   \\
	& \\
	- \frac{\bar{\psi}^1}{\bar{\psi}^2} & 1   \\   
	\end{array}
	\right) \left(
	\begin{array}{c}
	  \psi^1  \\
	  \\
	  \psi^2
	\end{array}
	\right) \, ,
	\eeq
or
	\beq
	\i \hbar
	\left(
	\begin{array}{c}
	  \dot{\psi}^1   \\
	  \\
	  \dot{\psi}^2  
	\end{array}
	\right)
	=
	\left(
	\begin{array}{ccc}
	H_1  &  V   \\
	& \\
	 \bar{V} &  H_2   \\   
	\end{array}
	\right)
	\left(
	\begin{array}{c}
	  \psi^1  \\
	  \\
	  \psi^2
	\end{array}
	\right)
	+ \frac{\gamma}{2} 
	\left(
	\begin{array}{ccc}
	-1/2  &  0  \\
	& \\
	-  \frac{\bar{\psi}^1}{\bar{\psi}^2} &  1/2   \\   
	\end{array}
	\right) \left(
	\begin{array}{c}
	  \psi^1  \\
	  \\
	  \psi^2
	\end{array}
	\right) \, .
	\eeq
It is not difficult to see that in both equations the normalization and the phase are not invariant and 
that they are involved in the evolution. This introduces nonlinearities in the equations, which as a consequence
are associated with non-Hermitian Hamiltonian operators. 

Finally, a major motivation for our work and for further investigation regards the applications of the formalism and a deeper comparison with
existing approaches. 
In the present work we have deliberately only scratched the surface of some
applications but we expect that, as it happened in the classical case, 
further systems may be analyzed from this perspective and possibly new tantalizing results will be found. 




\appendix
\section{Proof of Eq.~\eqref{Rho-Ev}}

\label{appA}

In this appendix we show that the contact master equation in Eq.~\eqref{eq:dotrho1} 
can be written {as} in Eq.~\eqref{Rho-Ev}. To do so, let us assume {that}
there exists an operator ${\mathbf A}({\bf z})$ of the form
	\beq \label{OpA}
	\mathbf{A}({\bf z}) = \left(
	\begin{array}{cc}
	\mathbb{A} & \mathbf{w}({\bf z}) \\
	&  \\
	\mathbf{w}^\dagger({\bf z}) & A({\bf z})
	  \end{array}
	\right) \,,
	\eeq
\noindent where $\mathbb{A}$ is an arbitrary $(n-1) \times (n-1)$-dimensional {Hermitian} matrix, 
$\mathbf{w}({\bf z})$ is {an} $(n-1)$-dimensional column vector and $A({\bf z})$ is a real number. 
Our goal is to prove that the anticommutator $\left[ \rho, {\mathbf A}({\bf z}) \right]_+$ of this operator and the density operator $\rho$ is given by

	\beq \label{A1}
	 \left[ \rho, {\mathbf A}({\bf z}) \right]_+ = 
	\frac{1}{1 + |{\bf z}|^2 } 
	\left(
	\begin{array}{cc}
	{\bf z} {\bf z}^\dagger 
	& \frac{{\bf z}}{2} \, (1 - |{\bf z}|^2) \\
	&  \\
	\frac{{\bf z}^\dagger}{2} \, (1 - |{\bf z}|^2)
	& - |{\bf z}|^2 
	  \end{array}
	\right) \, ,
	\eeq
\noindent where the density matrix is of the form
	\beq
	\rho 
	= \frac{1}{1 + |{\bf z}|^2 }
	\left(
	\begin{array}{ccc}
	{\bf z} {\bf z}^\dagger   & {\bf z} \\
	\\
	{\bf z}^\dagger    & 1 
	\end{array}
	\right) \, .
	\eeq
	
A direct computation of the anticommutator yields 
	\beq \label{Anticommutator}
	\left[ \rho, {\mathbf A}({\bf z}) \right]_+ = 
	\frac{1}{1 + |{\bf z}|^2 }
	\left(
	\begin{array}{ccc}
	\left[ {\bf z} {\bf z}^\dagger , {\mathbb A} \right]_+ 
	+ \mathbf{z}  \mathbf{w}^\dagger({\bf z}) + \mathbf{w}({\bf z}) \mathbf{z}^\dagger
	&& ( {\bf z} {\bf z}^\dagger + \mathbb{I})\, \mathbf{w}({\bf z}) + (\mathbb{A} + A({\bf z}) \, \mathbb{I} )\mathbf{z} \\
	\\
	\mathbf{w}^\dagger({\bf z}) ({\bf z} {\bf z}^\dagger + \mathbb{I})+ \mathbf{z}^\dagger( \mathbb{A}  + A({\bf z}) \, \mathbb{I})
	&& \mathbf{z}^\dagger \mathbf{w}({\bf z}) + \mathbf{w}^\dagger({\bf z}) \mathbf{z} + 2 A({\bf z}) 
	\end{array}
	\right) \, ,
	\eeq	
where $\mathbb{I}$ represents the $(n-1) \times (n-1)$-{dimensional} identity matrix. 
Equating  the right hand sides of~\eqref{A1} and~\eqref{Anticommutator} gives the following system of equations
 \begin{align}
	\left[ {\bf z} {\bf z}^\dagger , {\mathbb A} \right]_+ 
	+ \mathbf{z}  \mathbf{w}^\dagger({\bf z}) + \mathbf{w}({\bf z}) \mathbf{z}^\dagger
	&={\bf z} {\bf z}^\dagger \, , \label{Equation1} \\
	 ( {\bf z} {\bf z}^\dagger + \mathbb{I}) \mathbf{w}({\bf z}) +(\mathbb{A} + A({\bf z}) \, \mathbb{I} ) \mathbf{z}
	&=  \frac{{\bf z}}{2} \, (1 - |{\bf z}|^2) \, , \label{Equation21}\\
	\mathbf{w}^\dagger({\bf z}) ( {\bf z} {\bf z}^\dagger + \mathbb{I}) + \mathbf{z}^\dagger (\mathbb{A} + A({\bf z}) \, \mathbb{I} )
	&= \frac{{\bf z}^\dagger}{2} \, (1 - |{\bf z}|^2) , \label{Equation22}\\
	\mathbf{z}^\dagger \mathbf{w}({\bf z}) + \mathbf{w}^\dagger({\bf z}) \mathbf{z} + 2 A({\bf z}) 
	&=   - |{\bf z}|^2. \label{Equation3}
  \end{align}

Let us impose the conditions 
 \beq \label{CondAandA}
 \mathbb{A}^\dagger = \mathbb{A}, \qquad A^*({\bf z}) = A({\bf z}),
 \eeq
which guarantee that~\eqref{Equation22} is the conjugate transpose of~\eqref{Equation21}, i.e., Eq.~\eqref{Equation22} is no longer independent.
 
We now solve for $\mathbf{w}({\bf z})$ from Eq.~\eqref{Equation21} and obtain that
	\beq \label{Equationfora}
	\mathbf{w}({\bf z}) 
	= \left(  \mathbb{I} - \frac{\mathbf{z} \mathbf{z}^\dagger}{1 + |\mathbf{z}|^2}  \right)  
	\left( \frac{1}{2} (1 - |\mathbf{z}|^2) \,  \mathbb{I} - A({\bf z}) \,\mathbb{I} - \mathbb{A} \right) \mathbf{z},
	\eeq
where the inverse of the matrix $\mathbb{I} + \mathbf{z} \mathbf{z}^\dagger$ is given by 
	\begin{equation}
	\left(\mathbb{I} + \mathbf{z} \mathbf{z}^\dagger \right)^{-1} 
	=   \mathbb{I} - \frac{\mathbf{z} \mathbf{z}^\dagger}{1 + |\mathbf{z}|^2}  \, .
	\end{equation}
We then insert~\eqref{Equationfora} into~\eqref{Equation3} and solve for $A({\bf z})$ and obtain
	\beq \label{EquationforA}
	A({\bf z}) = \mathbf{z}^\dagger \, \mathbb{A}\, \mathbf{z} -  |{\bf z}|^2 \, .
	\eeq
Eq.~\eqref{EquationforA} allows us to express $\mathbf{w}({\bf z})$, given in~\eqref{Equationfora}, in terms of the matrix $\mathbb{A}$ as
	\beq \label{Equationfora2}
	\mathbf{w}({\bf z}) = \left(  \mathbb{I} - \frac{ \mathbf{z} \mathbf{z}^\dagger}{1 + |\mathbf{z}|^2} \right)  			\left( \frac{1}{2}(1 + |\mathbf{z}|^2) \, \mathbb{I} 
	- (\mathbf{z}^\dagger \, \mathbb{A} \, \mathbf{z}) \mathbb{I} - \mathbb{A} \right) \mathbf{z} \, .
	\eeq 
	
So far, we have obtained expressions for $A({\bf z})$ and $\mathbf{w}({\bf z})$, Eqs.~\eqref{EquationforA} and~\eqref{Equationfora2} respectively, 
	for which Eqs.~\eqref{Equation21},~\eqref{Equation22} and~\eqref{Equation3} are satisfied. 
The remaining equation~\eqref{Equation1} is trivially satisfied using the condition
	\beq
	\mathbf{z} \mathbf{w}^\dagger({\bf z}) + \mathbf{w}({\bf z}) \mathbf{z} 
	= - [\mathbf{z}\,\mathbf{z}^\dagger , \mathbb{A} ]_+ + \mathbf{z} \, \mathbf{z}^\dagger,
	\eeq
which, when inserted into~\eqref{Equation1} yields
	\beq
	\left[ { \bf z} {\bf z}^\dagger , \mathbb{A} \right]_+ 
	+ \mathbf{z} \mathbf{w}^\dagger({\bf z}) + \mathbf{w}({\bf z}) \mathbf{z}
	= \left[ { \bf z} {\bf z}^\dagger , \mathbb{A} \right]_+ 
	- [\mathbf{z}\,\mathbf{z}^\dagger , \mathbb{A} ]_+ 
	+\mathbf{z} \, \mathbf{z}^\dagger 
	= \mathbf{z} \, \mathbf{z}^\dagger .
	\eeq
As a result, Eq.~\eqref{Equation1} is trivially satisfied, without any further requirements on the Hermitian matrix $\mathbb{A}$.
Summing up, we have that the operator $\mathbf A({\bf z})$ in~\eqref{OpA}, with $\mathbb{A}$ being any Hermitian matrix, and $A({\bf z})$ and $\mathbf{w}({\bf z})$ 
given in~\eqref{EquationforA} and~\eqref{Equationfora2} respectively, 
satisfies~\eqref{A1}, thus concluding the proof.


\bibliographystyle{abbrvnat_mv}


\end{document}